\documentclass[12pt]{iopart}

\usepackage{iopams}
\usepackage{setstack}
\usepackage{graphicx}
\usepackage{subfigure}
\usepackage{color}

\newcommand{\be}{\begin{equation}}
\newcommand{\ee}{\end{equation}}
\newcommand{\bel}[1]{\begin{equation}\label{#1}}
\newcommand{\bea}{\begin{eqnarray}}
\newcommand{\eea}{\end{eqnarray}}
\newcommand{\ba}{\begin{array}}
\newcommand{\ea}{\end{array}}
\newcommand{\bef}{\begin{figure}}
\newcommand{\ef}{\end{figure}}

\begin{document}

\title[]{Nonlinear magneto-optical response to light carrying orbital angular momentum}

\author{Thomas Bose$^1$\footnote{Present address:
Department of Physics, University of York, Heslington,
York YO10 5DD, UK.} and Jamal Berakdar$^2$}
\address{$^1$ Theory Department, Max Planck Institute of Microstructure Physics, D-06120 Halle, Germany}
\address{$^2$ Institute of Physics, Martin-Luther-University, D-06120 Halle, Germany}

\begin{abstract}
We predict  a non-thermal magneto-optical effect for magnetic insulators subject to intense light
carrying orbital angular momentum (OAM).
Using a classical approach to
second harmonic generation in non-linear media with specific symmetry properties
we predict a significant nonlinear contribution to the local magnetic field triggered by light with OAM.
The resulting magnetic field originates from the displacement of electrons driven
by the electrical field (with amplitude $E_0$) of the spatially inhomogeneous optical pulse,
modeled here as a Laguerre-Gaussian beam carrying OAM.
In particular, the symmetry properties  of the irradiated magnet allow for magnetic field responses which are
second-order ($\sim E_0^2$) and fourth-order ($\sim E_0^4$) in electric-field strength
and have opposite signs.
For sufficiently high laser intensities, terms $\sim E_0^4$ dominate and
generate magnetic field strengths which can be as large as several Tesla.
Moreover, changing the OAM of the laser beam is shown to determine the direction
of the total light-induced magnetic field,
which is further utilized to study theoretically the non-thermal magnetization dynamics.

\end{abstract}


\maketitle

\section{Introduction}

The design and fabrication of ever smaller and faster magnetic devices for
data storage, sensorics and information processing
 entail the   development of efficient tools to control  the dynamic behavior of the magnetization.
In particular, femtosecond laser-induced magnetic excitations, originating in thermal and nonthermal effects,
offer the possibility to study magnetic systems on  time scales down to ($10-100\,\mathrm{fs}$) \cite{MFundNano:2006,Krilyuk:RevModPhys:p2731:2010}.
Thermal effects describe the energy transfer from a laser to the medium which is usually
thought to consist of three reservoirs - phonons, electrons and spins
\cite{Beaurepaire:PhysRevLett76:4250:1996,Kazantseva:EPL81:27004:2008,
Koopmans:PhysRevLett:85:844:2000,Vahaplar:PhysRevLett:103:117201:2009,
Radu:Nature472:205:2011,Ostler:NatComm:3:666:2012,Vahaplar:PhysRevB:85:104402:2012}, for a recent review see \cite{Krilyuk:RevModPhys:p2731:2010}.
Beside those thermal effects, nonthermal effects
can influence magnetization dynamics as well, e.g. the impulsive
stimulated Raman scattering \cite{Shen:NonlinearOptics:Book:1984} which was experimentally observed
\cite{Kalshnikova:PhysRevLett:99:167205:2007}
and theoretically described in the context of coherent magnon generation \cite{Gridnev:PhysRevB77:094426:2008}.
Furthermore, the inverse Faraday effect (IFE)
\cite{Pitaeevskii:SovPhysJETP:12:1008:1961,Ziel:PhysRevLett:15:190:1965,Pershan:PhysRev:143:574:1966,
Landau:ElecContMed:Book:1989} is another intriguing
nonthermal phenomenon
that refers to a build-up
of a static magnetization or an effective magnetic field in response
to circularly polarized light. On a microscopic level IFE relies heavily on the sample symmetry and/or
spin-orbital coupling (SOC).
Several experimental \cite{Kimel:Nature:435:2005,Hansteen:PhysRevB:73:014421:2006,
Steil:PhysRevB:84:224408:2011,Makino:PhysRevB:86:064403:2012,Mikhaylovskiy:PhysRevB:86:100405:100405,
Bakunov:PhysRevB:86:134405:2012}
as well as theoretical
\cite{Hertel:JMagMagMat:303:2006:L1,Zhang:JMagMagMat:321:L73:2009,Woodford:PhysRevB:79:212412:2009,
Yoshino:JMagMagMat:323:2531:2011,Taguchi:PhysRevB:84:174433:2011,Popova:PhysRevB:85:094419:2012,
Taguchi:PhysRevLett:109:127204:2012}
studies were performed during the recent past to obtain new insight into opto-magnetic
effects related to the IFE.
A microscopic description of the IFE often involves a Raman scattering process \cite{Shen:NonlinearOptics:Book:1984,Kimel:Nature:435:2005,Popova:PhysRevB:85:094419:2012}.
However, models were put forward to understand  IFE within the framework of classical single electron
currents \cite{Hertel:JMagMagMat:303:2006:L1,Zhang:JMagMagMat:321:L73:2009,Yoshino:JMagMagMat:323:2531:2011}.
Despite  the progress on opto-magnetic effects the role of higher order nonlinear processes
has not yet been addressed.
The aforementioned studies consider  effects characterized by a quadratic dependence of the optically generated magnetic field $B_{\rm ind}^{(1)}$ on the electric field of the light pulse $E_0$,
i.e. $B_{\rm ind}^{(1)}\propto E_0^2$.
It is one goal of the present paper to emphasize the role of higher order effects
in opto-magnetic phenomena, that is $B_{\rm ind}^{(2)}\propto E_0^4$.

In addition to the usage of laser beams to control the irradiated magnetic material the beam itself can be modified to achieve
a new form of magnetic sensing. E.g. one can engineer the time structure of the pulses such that orbital current is created \cite{Zhu_PRB_82_235304_2010,Matos_PRL94_166801_2005,Matos_EPL_69_277_2005,
Moskalenko_PRB_80_193407_2009,Moskalenko_PRB_74_161303_2006}.
The associated Oersted  field can then be utilized for steering magnetic dynamics  in the sample.
The other possibility is to shape appropriately the light pulses spatial structure,
for instance by using optical vortices, i.e. laser beams designed as to carry transferrable
orbital angular momentum (OAM) in
addition to the light polarization (associated with the photon spin)
\cite{Allen:PRA:45:8185:1992,Allen:ProgOpt:OAM:291:1999,Molina:NatPhys:3:305:2007}.

We note here that vortex beams of matter offer another promising opportunity for material research.
The generation of electron vortex beams was suggested theoretically in
\cite{Bliokh_PhysRevLett_99_190404_2007} and recently demonstrated experimentally by utilizing
spiral phase plates \cite{Uchida_Nature_464_737_2010}
and nanofabricated diffraction holograms \cite{Verbeeck_Nature_467_301_2010} achieving
vortex beams with high OAM numbers ($|l|=100$) \cite{McMorran_Science_331_192_2011}.
These works imply that electron vortices could lead to novel concepts in electron microscopy
providing additional information
about structure and properties of samples by analyzing the OAM-dependent signal.
Currently, experimental methods for the generation and modification of electron beams carrying OAM
are under intense research
\cite{Karimi:PRL:108:044801:2012, Schattschneider:PRL:109:084801:2012, Clark:PRL_111_064801:2013}
and electron vortex beams have even been
produced down to atomic resolution \cite{Verbeeck:APL:99:203109:2011}.
Likewise, the theoretical understanding of electron vortices was put forward by
investigating relativistic and nonparaxial corrections to
the scalar electron beams \cite{Bliokh:PRL_107_174802_2011}
and by studying the transfer of OAM from an electron vortex
to atomic electrons \cite{Lloyd_PRL_108_074802_2012, Yuan:PRA:88:031801:2013}.

In this present work we will investigate the effect of optical vortices, the theory however
is straightforwardly extendable
to the action of electron vortex beams, which have some similar as well as different characteristics
compared to optical beams, as discussed in
\cite{Bliokh_PhysRevLett_99_190404_2007} (the advantage of optical beams lies in their precise
temporal and frequency control while electron vortex beams are superior when it comes to spatial resolution).\\
Considering optical vortex beams it was predicted, based on a quantum mechanical theory,
that an OAM transfer for laser beams
interacting with molecules
will not occur between the light and the internal electronic-type motion in the electric dipole approximation,
but can take place
in the quadrupole interaction \cite{Babiker_PhysRevLett_89_143601_2002}.
Later on, this was experimentally verified in chiral matter
\cite{Araoka_PhysRevA_71_055401_2005,Loeffler_PhysRevA_83_065801_2011}.
In the present paper, we are particularly interested in the higher-order effect of the electronic motion
of the irradiated magnet
induced by laser light carrying OAM within a classical theory.

The core of our model is based on the classical anharmonic oscillator
\be
m_e\frac{d^2}{dt^2}x_\alpha(t)+2\tilde{\gamma}\frac{d}{dt}x_\alpha(t)+\frac{d}{dx_\alpha}V(\mathbf{x})=
F_\alpha(\mathbf{x},t) \,,
\label{NonlHarmOsc1}
\ee
with mass $m_e$, damping constant $\tilde{\gamma}$ and the potential $V(\mathbf{x})$ containing harmonic and
anharmonic parts, which is driven by the external force $\mathbf{F}(\mathbf{x},t)$.
This force results from the electrical field of the optical pulse and leads to a
time-dependent displacement $\mathbf{x}(t)$ (with components $x_\alpha$) of the electron which entails two effects.
On the one hand, an electric polarization is induced in the material and, on the other hand, microscopic electron
currents are generated which are responsible for the build-up of a magnetic field.
A theory based on Eq.~(\ref{NonlHarmOsc1}) with a harmonic potential $V(\mathbf{x})\propto \mathbf{x}^2$ describing the
latter process was presented in \cite{Yoshino:JMagMagMat:323:2531:2011} where the inverse Faraday effect
was mimicked by assuming a linear dependence between the electron displacement and the external electric field (note, however, that in these
works the sample symmetry effects were not considered, cf. in contrast the potential Eq.~(\ref{Potential}) used in this work).
Further effects are expected upon including non-linear response, e.g., second-harmonic generation (SHG).
 Thus, the electron motion is in general not only  affected by the fundamental frequency $\Omega$ of
the external field but also by the  SHG frequency $2\Omega$. This nonlinear effect is allowed in
all media lacking a
center of inversion symmetry \cite{Shen:NonlinearOptics:Book:1984,Boyd:NonlinearOptics:Book:2008}.
In the present study we consider thin magnetic garnet films which, although centrosymmetric in their bulk form,
can be used for SHG due to the loss of inversion symmetry at the surface and/or the presence
of a nonzero magnetization
\cite{Pan:PhysRevB:39:1229:1989,Pisarev:JPhysCondMat:5:8621:1993,Pavlov:PhysRevLett:78:2004:1997,
Kirilyuk:PhysRevB:61:R3796:2000,Gridnev:PhysRevB:63:184407:2001,Hansteen:PhysRevB70:094408:2004}.

We consider circularly polarized laser pulses which are spatially inhomogeneous, and
additionally, carry OAM.
More explicitly, we employ cylindrically symmetric Laguerre-Gaussian beams \cite{Allen:PRA:45:8185:1992}
which are able to produce local orbital currents (e.g., \cite{Koeksal:PhysRevA:86:063812:2012} and references therein) and hence orbital
local magnetic fields that can be utilized for initiating locally spin dynamics in the sample.
The influence of light with OAM on the spin-degrees of freedom in a condensed matter
system was discussed in \cite{Quinteiro_OptExpr_19_26733_2011}, here this aspect is not considered in our
 classical treatment. The time
structure of these magnetic field pulses is related mainly to the laser pulse duration. As shown below, substantial intensities
are needed. Hence, short pulses should be used to minimize radiation damage.

The paper is organized as follows: In Sec.~\ref{Sec_Theory} we introduce the theoretical model emphasizing the
role of the geometry and crystallographic and magnetization-induced symmetry under consideration. Further, we
explain the procedure applied to calculate the time-dependent light-induced magnetic field. This induced
magnetic field is then used to model magnetization dynamics.
After the detailed description of the model parameters we present and discuss our results in
Sec.~\ref{Sec_Results}.
Finally, the results are summarized and the corresponding conclusion is given in Sec.~\ref{Sec_Conclusion}.

\section{Theory\label{Sec_Theory}}

We investigate the physical situation sketched in Fig.~\ref{Fig_Scheme}.
\bef
\centering \includegraphics[width=7.5cm]{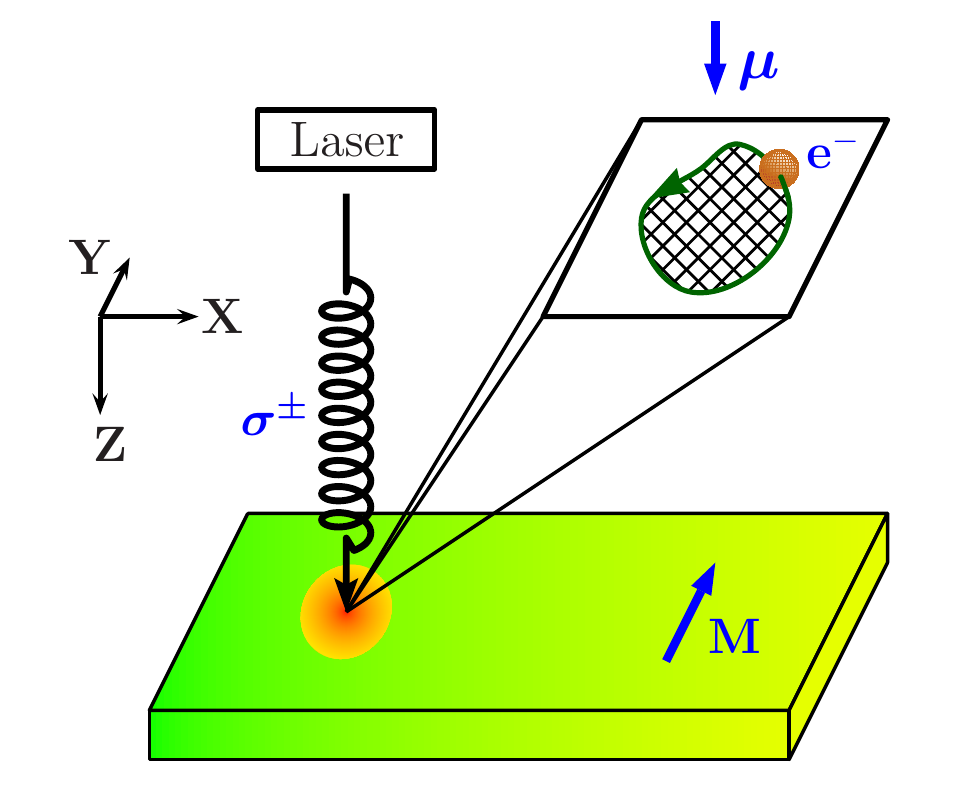}
\caption{(Color online) Schematic of the system. $\mathbf{M}$ is the magnetization
of the sample, ${\rm e}^-$ indicates the single-electron current, $\boldsymbol\mu$ designates
the magnetic moment of
the current loop and $\sigma^\pm$ is the polarization of the circularly polarized light pulse,
(right-handed: $+$, left-handed: $-$). Further description is given in the text.}
\label{Fig_Scheme}
\ef
An optical pulse is applied to a magnetic sample with magnetization $\mathbf{M}$
in $\mathrm{\hat {e}}_{Y}$-direction.
As a consequence of the circularly polarized light
microscopic electronic currents will be generated. Moving on a closed curve the electron current generates
a magnetic moment $\boldsymbol\mu$.
The absolute value of the magnetic moment is given by the area enclosed by the
current loop $\Lambda$ multiplied by the strength of the single-electron current $j_e$,
i.e. $|\boldsymbol\mu|=j_e\,\Lambda$.
Due to the inhomogeneity of the laser pulse the absolute value of the magnetic moment differs spatially.

We are interested in the two-dimensional motion of the electrons in the film plane ($e_{z}$ is along the film normal).
For that purpose we refer to different coordinate systems as shown in Fig.~\ref{FigCoordinate}.
The laser pulse impinges on the surface of the magnetic film and generates an intensity profile which varies in
the $\mathrm{\hat {e}}_{X}$-$\mathrm{\hat {e}}_{Y}$-plane. The origin of the $\mathrm{\hat {e}}_{X}$-$\mathrm{\hat {e}}_{Y}$-coordinate system coincides
with the center of the light beam.
We will operate in cylindrical coordinates with $e_{z}$  coinciding with the light propagation 
direction ($(r,\phi)$ indicate the planar spatial position),
and assume the electron initial velocity distribution (given by the Compton profile) to be 
subsidiary with the respect to the
intense laser induced velocities.
%
Applying the laser pulse the electrical
field of the optical pulse couples to the electronic charge which leads to a displacement of the electron.
This motion of the electron is described in the $\mathrm{x}$-$\mathrm{y}$ (displacement) frame.
Comparing the typical length scale of the laser pulse ($\lambda_L\sim \mu \mathrm{m}$) with that of
the induced electron displacement ($\lambda_D \lesssim \mathrm{nm}$) it follows
that $\lambda_D\ll \lambda_L$.
Therefore when treating the charge dynamics, the spatial variation of the electrical field of the laser pulse in the
$\mathrm{x}$-$\mathrm{y}$-frame (i.e., in the \emph{displacement frame}) may be neglected.
\bef
\centering \includegraphics[width=7.5cm]{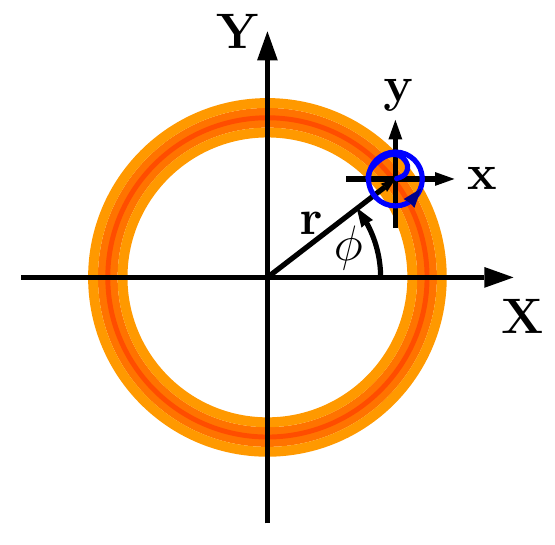}
\caption{(Color online) Coordinate systems considered. The spatial inhomogeneous laser pulse produces an intensity profile in the $\mathrm{\hat {e}}_{X}$-$\mathrm{\hat {e}}_{Y}$-plane. Here, only the region of highest intensity is shown. The electron is initially located at the origin of the $\mathrm{x}$-$\mathrm{y}$-frame at position
$(r,\phi)$ according to the center of the light beam. Due to the electrical field of the laser pulse the
electron starts to move.}
\label{FigCoordinate}
\ef

The laser light is described by a circularly polarized Laguerre-Gaussian beam propagating
in $\mathrm{Z}$-direction. Using cylindrical coordinates $r=\sqrt{X^2+Y^2}$, $\phi=\arctan[Y/X]$ and $Z=Z$
the corresponding electrical field distribution is given by
\bea
\mathbf{E}(r, \phi, Z;t) = \mathbf{E}_0(r,\phi,Z)\,\exp[-\mathrm{i}\tilde{\Omega}t]
+ \mathrm{c.c.} \,,\nonumber \\
\mathbf{E}_0(r,\phi,Z) = 2\hat e_\sigma E_0\,\mathcal{F}_l^p(r,\phi)\,\exp[\mathrm{i}q_ZZ]\,,\nonumber \\
\tilde{\Omega} = \Omega -\frac{\mathrm{i}}{\tau_p} \,,
\label{ElFieldLGbeam}
\eea
with the fundamental frequency $\Omega$ and the field pulse duration $\tau_{\rm p}$.
$E_0$ is the amplitude of the electrical field, $q_Z$ is the wave vector along the $\mathrm{Z}$-axis,
$\hat e_\sigma=\frac 1 2 (\mathbf{\hat {e}}_{X}+\mathrm{i}\,\sigma\mathbf{\hat {e}}_{Y})$ is the vector of the light polarization (related to the photon spin) with
$\sigma=\pm 1$ standing for right-handed ($+1$),  or left-handed ($-1$) helicity;
and ${\rm c.c.}$ stands for the complex conjugate of the first term.
The function $\mathcal{F}_l^p$ describes the spatial structure of the light beam and has the form
\be
\mathcal{F}_l^p(r,\phi) = \frac{1}{N_l}\, \mathcal{L}_p^{|l|}\left( \frac{2\,r^2}{w_0^2}\right)\,\exp\left[ \frac{-r^2}{w_0^2}\right]
\,\left( \frac{\sqrt{2}\,r}{w_0}\right)^{|l|}\,\exp[\mathrm{i}\,l\phi] \,,
\label{Eq_F_lp}
\ee
with the associated Laguerre polynomial $\mathcal{L}_p^{|l|}$.
The parameter $l \in \mathbb{Z}$ is the topological charge of the optical vortex and can be considered as the
amount of OAM transferred to the charge when interacting with the laser beam.
We note that in case of $l\neq 0$, Eq.~(\ref{Eq_F_lp}) describes an optical vortex with topological charge $l$.
For later purposes we introduce the absolute value of the OAM number as $L=|l|$.
Further parameters are the number of radial nodes $p$, the waist of the beam $w_0$
and the normalization constant $N_l$ which depends on $l$. 
The latter is introduced to ensure that
the laser power transferred to the medium is independent of the OAM represented by $l$.

\subsection{Electron displacement}

Due to the presence of anharmonicity we expand the electron displacement coordinates $x_\alpha =x,y$
according to
\be
x_\alpha=\epsilon\,x_\alpha^{(1)} + \epsilon^2\,x_\alpha^{(2)} + \epsilon^3\,x_\alpha^{(3)} + ... \,\,,
\label{DisplExpansion}
\ee
with the small perturbation parameter $\epsilon$ (which will be set equal to one later).
It is our aim to analyze the effect of second harmonics and, therefore, we calculate the solution up to the
order $\epsilon^2$.
The new aspect in this work is that we consider the potential to consist of harmonic and anharmonic contributions, $V_{\rm h}$ and $V_{\rm anh}$,
and has the form
\be
V(\mathbf{x})= V_{\rm h}(\mathbf{x}) + V_{\rm anh}(\mathbf{x})
      = \frac{m_e}{2}\,\omega_{0\alpha}^2\,x_\alpha^2 + \frac{m_e\,a}{3}\,\Gamma_{\alpha\beta\gamma} x_\alpha x_\beta x_\gamma\,,
\label{Potential}
\ee
where  sums over repeated  indices is implicitly assumed.
The eigenfrequencies are given by $\omega_{0\alpha}$
and the anharmonic constant $a$.
$\Gamma_{\alpha\beta\gamma}$ is the higher order coupling tensor.
The potential in Eq.~(\ref{Potential}) is an appropriate choice for noncentrosymmetric media \cite{Boyd:NonlinearOptics:Book:2008}.

The evolution equation (\ref{NonlHarmOsc1}) for the first-order displacement $\propto \epsilon$
which is driven by the
external electric field given in Eqs.~(\ref{ElFieldLGbeam}) and (\ref{Eq_F_lp}) is cast as
\be
\frac{d^2}{dt^2}x_\alpha^{(1)}+2\gamma\frac{d}{dt}x_\alpha^{(1)}+\omega_{0\alpha}^2\,x_\alpha^{(1)}
=-\frac{e}{m_e}\,E_\alpha(r,\phi,t) \,,
\label{xDetOrdnung1}
\ee
where we have substituted $\tilde{\gamma}/m_e=\gamma$.

We are interested in the solution of Eq.~(\ref{xDetOrdnung1}) which follows the electrical field of
the optical pulse, that is we neglect the fast decaying transient solutions.
Assuming $\omega_{0x}^2=\omega_{0y}^2=\omega_{0}^2$ and $Z=0$, the displacement of the
electron in the $\mathrm{x}$-$\mathrm{y}$-frame becomes
\bea
&x^{(1)}(t) = -\frac{e\,E_0}{m_e}\,F_l^p(r,\phi)\,\frac{\exp\left[-{\rm i}\tilde{\Omega}t\right]}{\psi\left(\tilde{\Omega}\right)} \,,\nonumber \\
&x_{\rm tot}^{(1)}(t) = x^{(1)}(t) + x^{(1)*}(t) \,,\nonumber \\
&y_{\rm tot}^{(1)}(t) = y^{(1)}(t) + y^{(1)*}(t) = \sigma{\rm i}\,\left(x^{(1)}(t) - x^{(1)*}(t)\right) \,,
\label{xDetOrdnung1Sol}
\eea
which is the particular solution of Eq.~(\ref{xDetOrdnung1}).
We note that the expression for the total displacement $\left(x_{\rm tot}^{(1)},\,y_{\rm tot}^{(1)}\right)$
in Eq.~(\ref{xDetOrdnung1Sol}) yields real quantities.
The function $\psi\left(\tilde{\Omega}\right)$ reads
\be
\psi\left(\tilde{\Omega}\right) =
\omega_{0}^2 - \tilde{\Omega}^2 - 2{\rm i}\gamma \tilde{\Omega} \,,
\label{PsiVonOmega1}
\ee
with $\tilde{\Omega}$ given in Eq.~(\ref{ElFieldLGbeam}).

The equation of motion of the second-order electron displacements is
\bea
\frac{d^2}{dt^2}x_\mu^{(2)}+2\gamma\frac{d}{dt}x_\mu^{(2)}+\omega_0^2\,x_\mu^{(2)}
= -\frac{1}{m_e}\frac{d V_{\rm anh}(\mathbf{x})}{dx_\mu} \,,\nonumber \\
\frac{d V_{\rm anh}(\mathbf{x})}{dx_\mu} = \,\frac{m_e a}{3} \,\Gamma_{\alpha\beta\gamma\delta}\,
\left(x_\beta x_\gamma M_\delta\,\delta_{\alpha\mu}
 + x_\alpha x_\gamma M_\delta\,\delta_{\beta\mu}
+ x_\alpha x_\beta M_\delta\,\delta_{\gamma\mu}\right) \,,
\label{xDetOrdnung2}
\eea
with the  Kronecker delta symbol $\delta_{\rho\nu}$. The anharmonic potential
$V_{\rm anh}$ is given in Eq.~(\ref{Potential}).
An important statement of Eq.~(\ref{xDetOrdnung2}) is that the actual symmetry properties of the
material do matter.
To obtain an explicit solution the nonzero components of the coupling tensor $\Gamma_{\alpha\beta\gamma}$
have to be determined first. $\Gamma_{\alpha\beta\gamma}$ is related to the second-order
susceptibility $\chi_{\alpha\beta\gamma}$ via the expression
\be
\mathcal{P}^{(2)}_\alpha = \epsilon_0\,\chi_{\alpha\beta\gamma}E_\beta E_\gamma =e\,N_e\,x_\alpha^{(2)}
\label{GammaChi1}
\ee
for the components of the second-order nonlinear electric polarization $\mathcal{P}^{(2)}_\alpha$.
Here $x_\alpha^{(2)}$ are the components
$\left(x^{(2)},\,y^{(2)}\right)$ of the second-order displacement, and $\epsilon_0$ and $N_e$ are
the permittivity of free space and the number of oscillators per unit volume.
Therefore, the symmetry properties of the susceptibility $\chi_{\alpha\beta\gamma}$
can be transferred to the coupling tensor $\Gamma_{\alpha\beta\gamma}$.
For thin magnetic garnet films, which were shown to be of great importance for the investigation
of nonthermal effects \cite{Hansteen:PhysRevLett:95:047402:2005,Hansteen:PhysRevB:73:014421:2006},
the second-order susceptibility comprises two terms, a crystallographic contribution
$\chi_{\alpha\beta\gamma}^{\rm cr}$ and a magnetic contribution
$\chi_{\alpha\beta\gamma\delta}^{\rm m}\,M_\delta$
\cite{Pavlov:PhysRevLett:78:2004:1997}, where $M_\delta$ are the components of the magnetization in the
sample frame.
For the geometry depicted in Fig.~\ref{Fig_Scheme} we assume (001)-oriented films. Thus, the corresponding point group
symmetry is $4mm$ ($C_{4\nu}$) and the crystallographic contribution vanishes for normal incidence of the optical pulse
\cite{Pavlov:PhysRevLett:78:2004:1997,Gridnev:PhysRevB:63:184407:2001}.
Hence, the SHG is purely magnetization induced and the nonvanishing components of
$\chi_{\alpha\beta\gamma\delta}^{{\rm m}}$ are
$\chi_{xxxy}^{{\rm m}} = -\chi_{yyyx}^{{\rm m}}$,
$\chi_{xxyx}^{{\rm m}} = \chi_{xyxx}^{{\rm m}} = -\chi_{yxyy}^{{\rm m}} = -\chi_{yyxy}^{{\rm m}}$, and
$\chi_{yxxx}^{{\rm m}} = -\chi_{xyyy}^{{\rm m}}$,
see \cite{Gridnev:PhysRevB:63:184407:2001}.
Assuming the same symmetry for the coupling tensor
$\Gamma_{\alpha\beta\gamma} = \Gamma_{\alpha\beta\gamma\delta}^{{\rm m}}\,M_\delta$,
its nonvanishing components are
\bea
\Sigma_1 = \Gamma_{xxxy}^{{\rm m}} &= -\Gamma_{yyyx}^{{\rm m}} \,, \nonumber \\
\Sigma_2 = \Gamma_{xxyx}^{{\rm m}} = \Gamma_{xyxx}^{{\rm m}} &= -\Gamma_{yxyy}^{{\rm m}}
= -\Gamma_{yyxy}^{{\rm m}}\,, \nonumber \\
\Sigma_3 = \Gamma_{yxxx}^{{\rm m}} &= -\Gamma_{xyyy}^{{\rm m}} \,.
\label{NonzeroComponents1}
\eea
For a magnetic garnet film with magnetization in $\mathrm{y}$-direction, i.e. $\mathbf{M}=M_s\,\mathbf{\hat {e}}_{Y}=M_s\,\mathbf{y}$,
after some algebra we find the evolution equation for the second order
displacement $\propto \epsilon^2$ which has the form
\bea
\fl \frac{d^2}{dt^2}x^{(2)}+2\gamma\frac{d}{dt}x^{(2)}+\omega_0^2\,x^{(2)} =&
-a\,M_s\,\left[(3\Sigma_1+(2\Sigma_2+\Sigma_3))\left(x^{(1)^2}+x^{(1)*^2}\right) \right. \nonumber \\
&\left.+ 2\,(3\Sigma_1-(2\Sigma_2+\Sigma_3))\,x^{(1)}x^{(1)*}\right] \nonumber \\
\fl \frac{d^2}{dt^2}y^{(2)}+2\gamma\frac{d}{dt}y^{(2)}+\omega_0^2\,y^{(2)} =&
\sigma{\rm i}\,2\,aM_s\,(2\Sigma_2+\Sigma_3)\left( x^{(1)^2}-x^{(1)*^2}\right)\,.
\label{EOM_2ndOrder}
\eea
The expression for $x^{(1)}$ is given in Eq.~(\ref{xDetOrdnung1Sol}).
Thus, the particular solution of Eq.~(\ref{EOM_2ndOrder}) can be expressed as
\bea
\fl x_{\rm tot}^{(2)}(t) = x^{(2)}(+2\Omega,+\Omega,+\Omega;\,t)
+ x^{(2)}(-2\Omega,-\Omega,-\Omega;\,t) + x^{(2)}(0,+\Omega,-\Omega;\,t) \,, \nonumber\\
\fl y_{\rm tot}^{(2)}(t) = y^{(2)}(+2\Omega,+\Omega,+\Omega;\,t) + y^{(2)}(-2\Omega,-\Omega,-\Omega;\,t)
\label{xDetOrdnung2Sol1}
\eea
where the $\pm 2\Omega$ terms represent the second harmonics and the last term in the expression
for $x_{\rm tot}^{(2)}(t)$ is usually referred to as optical rectification
\cite{Bass:PhysRevLett:9:446:1962,Shen:NonlinearOptics:Book:1984,Boyd:NonlinearOptics:Book:2008}.
Therefore, a dc-shift of the electron is only observed in the $\mathrm{x}$-direction which is
a direct consequence of the underlying, magnetization-induced, symmetry.

\subsection{Optically-generated magnetic field\label{Sec_Opt_Gen_Field}}

Solving the evolution equations for the first-order and second-order displacement of the electrons
derived in the preceding section enables the calculation of the optically-generated magnetic field.
Illuminating the magnetic film with circularly polarized light forces the electron to move on a loop
which is closed for laser pulses of infinite length.
For a finite pulse width $\tau_p$ the electron trajectory can be approximated by a closed loop
in case $\tau_p^{-1}\ll\Omega$.
This condition is approximately fulfilled in our model.
Calculating the trajectory of the single-electron allows to determine the magnetic moment associated with
this displacement. The total optically-induced magnetic field is obtained by summing over all
contributing electron orbits.
The magnetic moment $\mu$ of a closed current loop of a single-electron is given by $j_e\,\Lambda(r,\phi;t)$,
where $j_e$ is the single-electron current and $\Lambda(r,\phi;t)$ is the time dependent area
enclosed by the current loop in the coordinate system of the light beam.
Separating first-order and second-order displacements the total light-induced magnetic moment is then given by
\bea
\boldsymbol\mu^{\rm tot}(r,\phi;t) &= \boldsymbol\mu^{(1)}(r,\phi;t) + \boldsymbol\mu^{(2)}(r,\phi;t) \,, \nonumber \\
\mu^{(1)}(r,\phi;t) &= \frac{e\,\Omega}{2\,\pi}\,\Lambda^{(1)}(r,\phi;t) \,,\nonumber \\
\mu^{(2)}(r,\phi;t) &= \frac{e\,\Omega}{\pi}\,\Lambda^{(2)}(r,\phi;t) \,.
\label{MagMoment2}
\eea
Here, we introduced the areas $\Lambda^{(1)}$ and $\Lambda^{(2)}$ enclosed by the first-order and
second-order electronic displacements.
Right-handed (left-handed) circularly polarized beam creates a light-induced magnetic moment $\boldsymbol\mu^{(1)}$ in
$+Z$ ($-Z$)-direction. The direction of the second-order magnetic moment $\boldsymbol\mu^{(2)}$ strongly
depends on the
coupling tensor $\Gamma_{\alpha\beta\gamma}$ introduced in Eq.~(\ref{NonzeroComponents1}).
The total light induced magnetic field in $\mathrm{Z}$-direction can then be calculated from
\be
\mathbf{B}_{\rm ind}(t)=N_e\, \mu_0\,\frac{1}{\Lambda_0}\,\int\limits_0^{R_0}\boldsymbol\mu^{\rm tot}(r,\phi;t)\,\,r\,dr\,d\phi \,,
\label{MagField}
\ee
where $N_e$ and $\mu_0$ are the electron number density and the magnetic permeability.
$\Lambda_0=\pi\,R_0^2$ is the reference area over which the laser pulse influences
the magnetic structure of the film.

The details of the procedure are as follows:
First, we solved the equations of motion of the first-order and second-order single-electron displacements,
Eqs.~(\ref{xDetOrdnung1}) and (\ref{EOM_2ndOrder}),
for variable  distance $r$ between the electron and the center of the light-beam, i.e. the origin
of the $\mathrm{\hat {e}}_{X}$-$\mathrm{\hat {e}}_{Y}$-coordinate system.
The radial distance was varied in the range $0\leq r\leq 5\,\mu\mathrm{m}$ with a resolution of
$\Delta r = 2.5\,\mathrm{nm}$.
For each $r$ the time-dependent single-electron displacement was determined.
The peak field of the laser pulse sets in at $t=0$ and diminishes exponentially within the time $\tau_p$ (cf. Eq.(\ref{ElFieldLGbeam})).
A waiting time of one cycle period $2\pi/\Omega\simeq 2.7\,\mathrm{fs}$ until the
transient solutions decreased significantly was taken into account.
After the second full cycle of the electron, i.e. after $5.4\,\mathrm{fs}$, the value of the
light-induced magnetic field was calculated by computing the area enclosed by the electron trajectory.
We point out that the elliptic motion of the electron according to the second-order displacement
was already performed twice due to the motion with frequency $2\Omega$.
In the same manner the magnetic moment of the single-electron was calculated each time the
electron completes a full circle.
In total, $29$ values of the time-dependent magnetic moments according to Eq.~(\ref{MagMoment2})
were computed; the last value corresponding to the time $81\,\mathrm{fs}$ after the laser
pulse was applied.
Finally, the average total magnetic moment and the total optically-induced magnetic
field were calculated according to Eq.~(\ref{MagField}).

\subsection{Magnetization dynamics}

The light-induced magnetic field in Eq.~(\ref{MagField}) is used to excite magnetization dynamics
described by the Landau-Lifshitz-Gilbert equation \cite{Landau:ZdS:8:p153:1935,Gilbert:ITOM:40:p3443:2004}
\be
\frac{\partial \mathbf{m}}{\partial t}=-\frac{\gamma_e}{1+\alpha^2}\,\left(\mathbf{m}\times \mathbf{B}_{\rm eff}
+ \alpha\,\mathbf{m}\times \left[\mathbf{m}\times \mathbf{B}_{\rm eff}\right]\right) \,,
\label{LLGeq}
\ee
with the damping parameter $\alpha$, the absolute value of the gyromagnetic
ratio for electrons $\gamma_e$ and the effective magnetic field $\mathbf{B}_{\rm eff}$.
The magnetization vector $\mathbf{m}$ is a unit vector according to $\mathbf{m}=\mathbf{M}/M_s$,
where $M_s$ is the saturation magnetization.
In general, the effective field consists of different contributions, such as static magnetic fields,
demagnetization fields, anisotropy fields and exchange fields.
However, in the present study we consider the case when the static magnetic field $\mathbf{B}_0$ is supplemented
by the light induced magnetic field $\mathbf{B}_{\rm ind}(t)$, i.e.
\be
\mathbf{B}_{\rm eff}(t)=\mathbf{B}_0 + \mathbf{B}_{\rm ind}(t) \,.
\label{Beff}
\ee
Utilizing this effective magnetic field we solved Eq.~(\ref{LLGeq}) numerically.

\subsection{Model parameters}

Nonlinear magneto-optical phenomena in magnetic garnet films
involving second harmonics generation, which are not related to optically-induced magnetic fields,
were studied experimentally in
\cite{Pan:PhysRevB:39:1229:1989,Pisarev:JPhysCondMat:5:8621:1993,Pavlov:PhysRevLett:78:2004:1997,
Kirilyuk:PhysRevB:61:R3796:2000,Gridnev:PhysRevB:63:184407:2001,Hansteen:PhysRevB70:094408:2004}.
Regarding intensities, in \cite{Gridnev:PhysRevB:63:184407:2001} the authors report an average power
density of approximately $10^{3}\,\mathrm{W}/\mathrm{cm}^2$.
Whereas, in \cite{Hansteen:PhysRevB:73:014421:2006}, where the nonthermal optical control of the
magnetization in magnetic garnet films was studied, a laser peak power density of
about $10^{11}\,\mathrm{W}/\mathrm{cm}^2$ was applied by $100\,\mathrm{fs}$ laser pulses.
In the present theoretical study both aforementioned values do not lead to the generation of
significant second-order effects, i.e. the build-up of a magnetic field induced
by the second order electron displacement, which is comparable to the effects of first order.
To observe first-order and second-order effects ranging in similar orders of magnitude we have chosen
peak intensities
between $2\cdot 10^{14}- 2\cdot 10^{16}\,\mathrm{W}/\mathrm{cm}^2$.
Taking into account a reference radius of $R_0=5\,\mu\mathrm{m}$ and a pulse
duration of $\tau_p=10\,\mathrm{fs}$, as applied in the present model,
pump pulse energies of $1.6\,\mu\mathrm{J}$ up to $160\,\mu\mathrm{J}$ are required.
The above mentioned intensities are achieved by varying the electrical field strength
in the interval
$1.85\cdot 10^{8}\,\mathrm{V}/\mathrm{cm}\leq E_0\leq 1.85\cdot 10^{9}\,\mathrm{V}/\mathrm{cm}$.
In particular, we will make use of the reference intensity
$I_0=2\cdot 10^{14}\,\mathrm{W}/\mathrm{cm}^2$ calculated at $t=0$ for the electrical field
strength of $1.85\cdot 10^{8}\,\mathrm{V}/\mathrm{cm}$.
The range of the absolute values of the topological charge (OAM) was chosen to be $0\leq L=|l|\leq 6$.

The spatial distribution of the intensities for different values of $L$ is
shown in Fig.~\ref{Fig_Intensity_2} (for simplicity we set the radial node to be $n=0$).
\bef[h!]
\centering
\subfigure[$L=0$]{
\includegraphics[width=6cm]{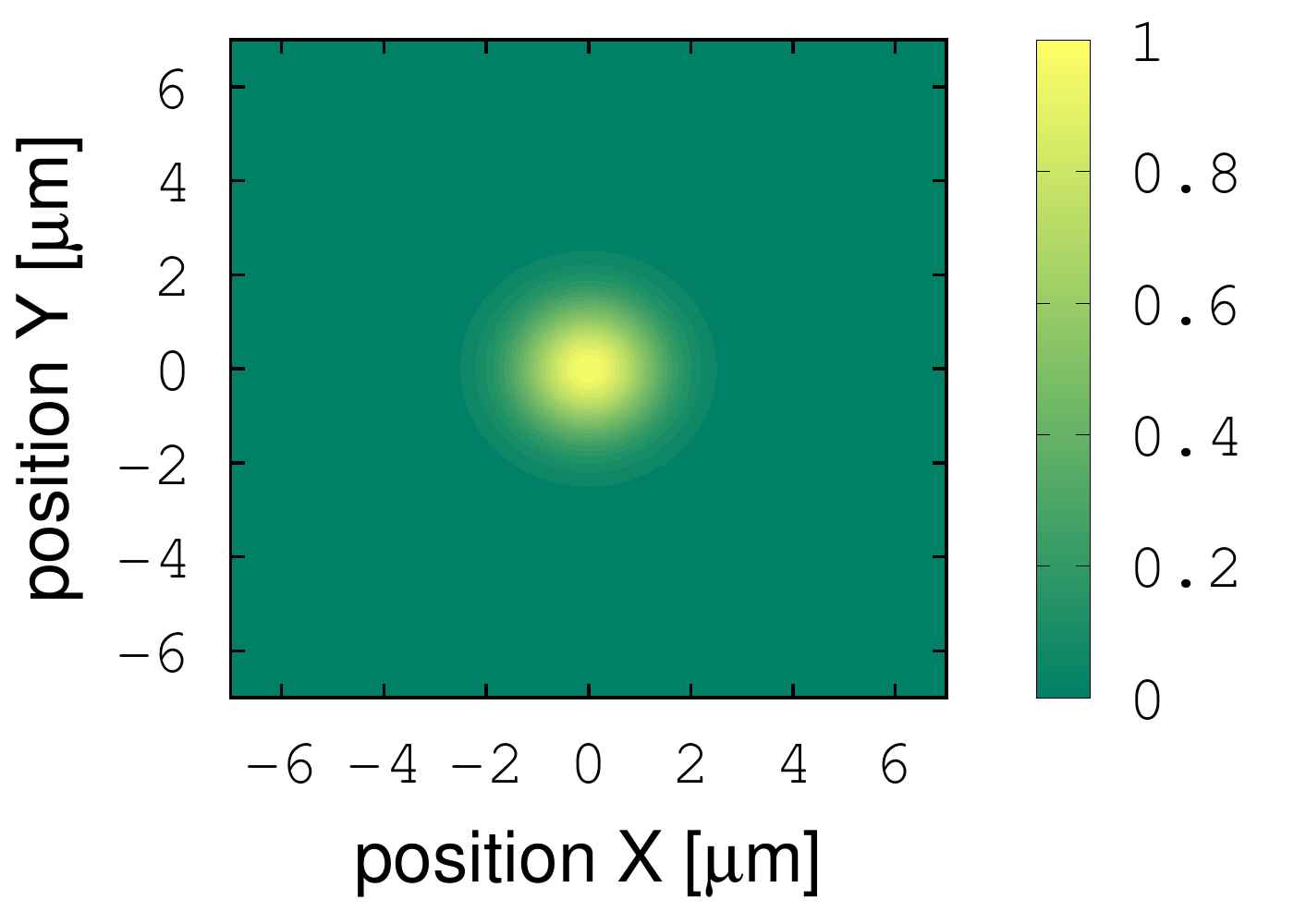}}
\subfigure[$L=2$]{
\includegraphics[width=6cm]{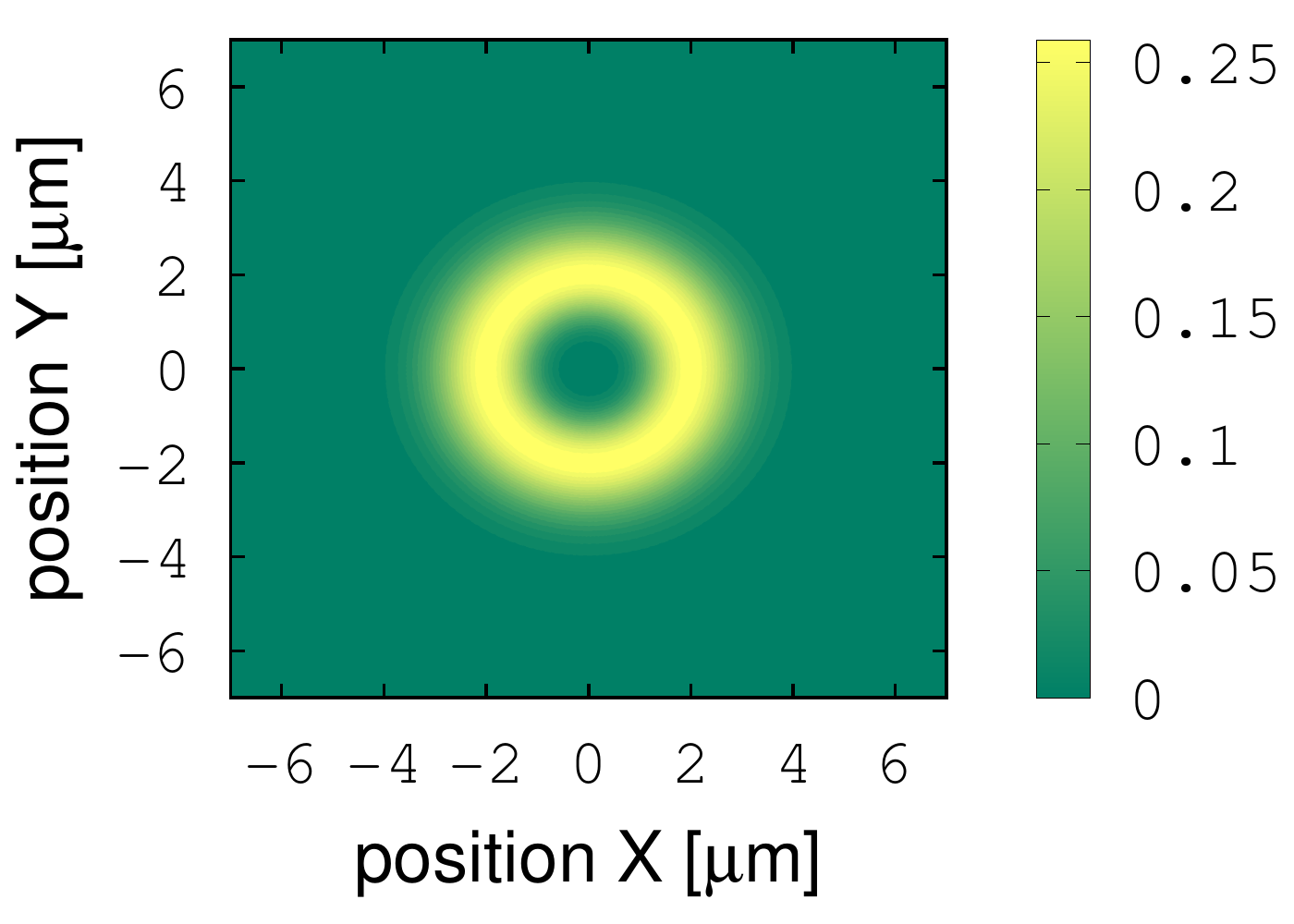}}\\
\subfigure[$L=4$]{
\includegraphics[width=6cm]{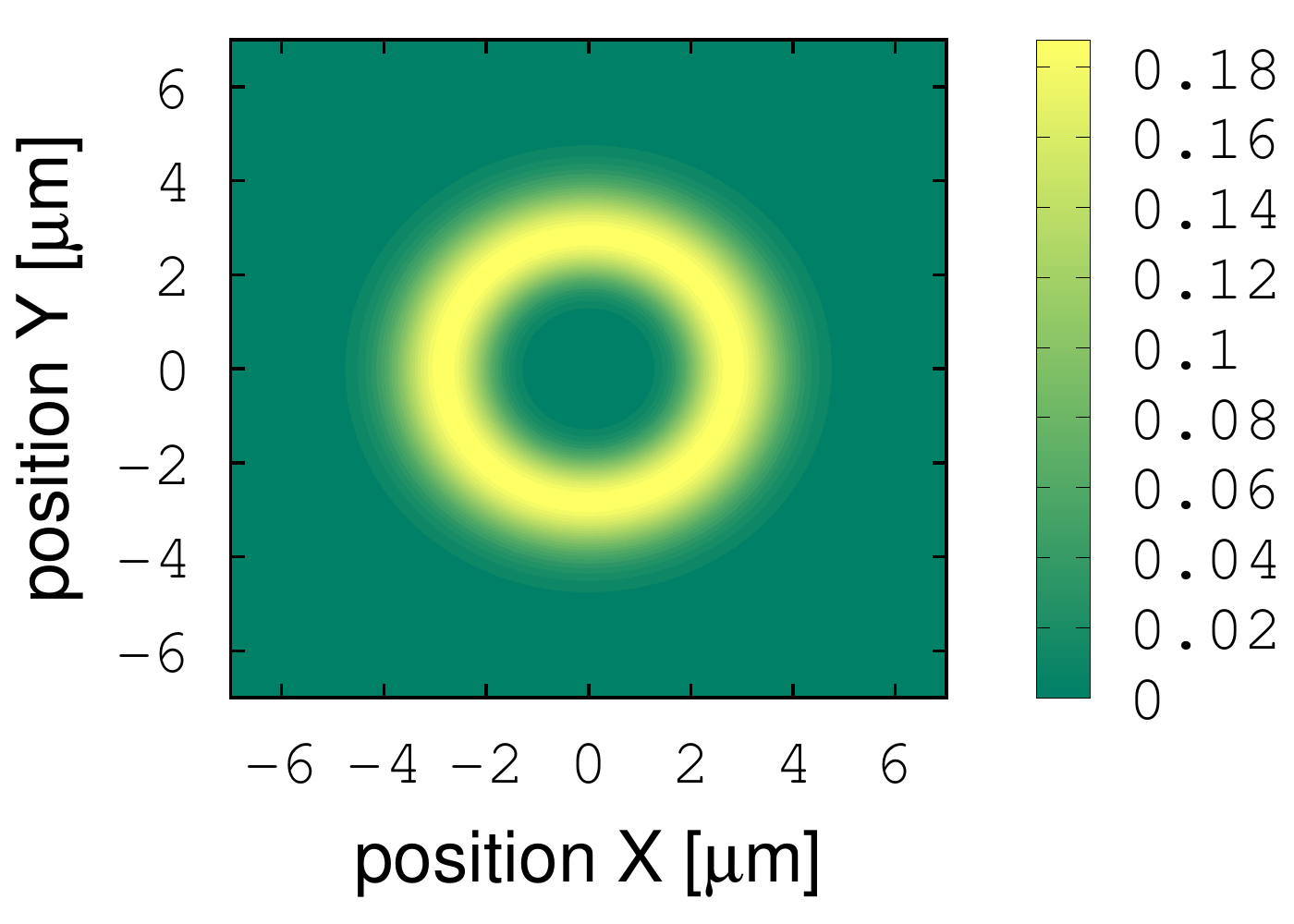}}
\subfigure[$L=6$]{
\includegraphics[width=6cm]{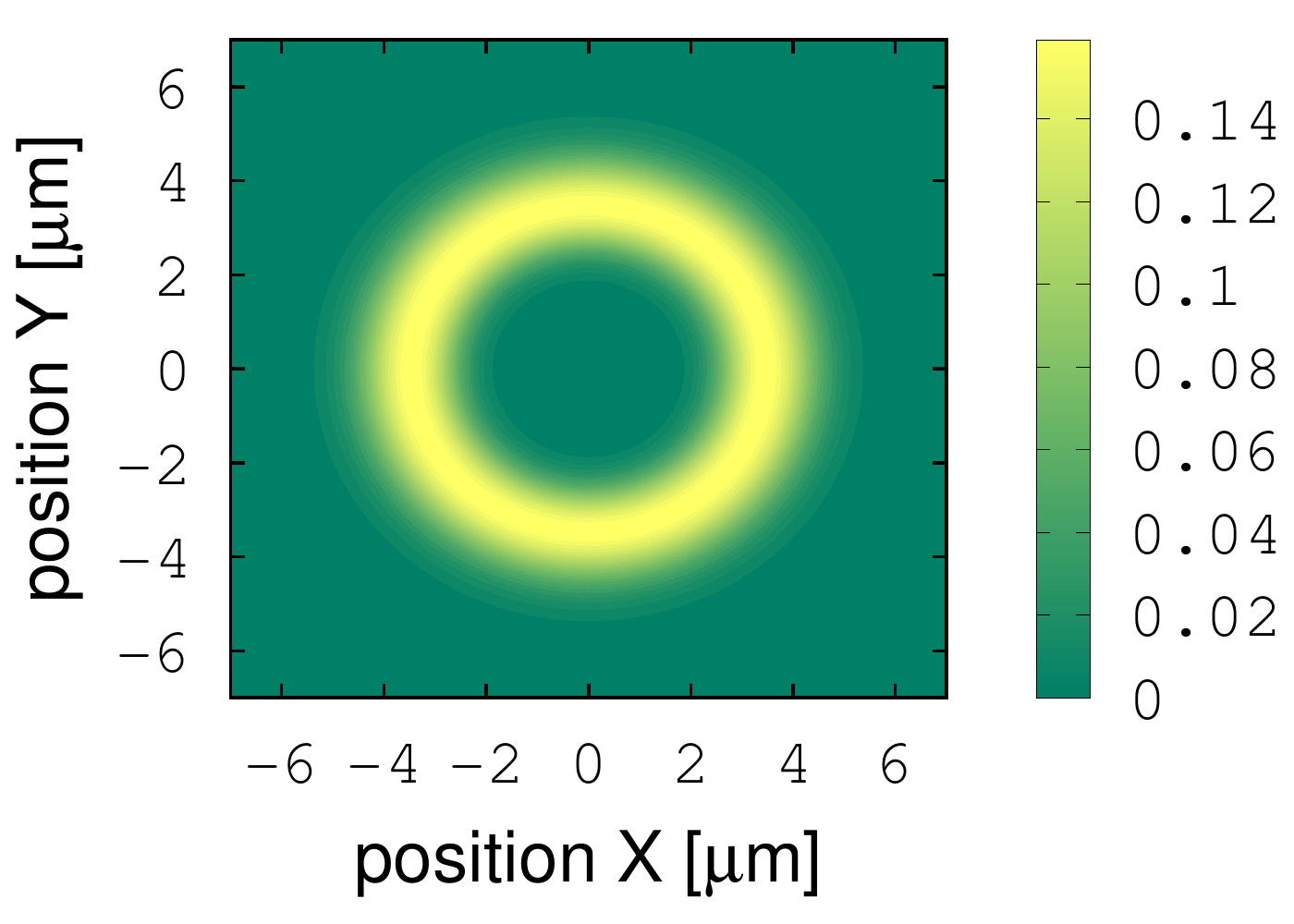}}
\caption{\label{Fig_Intensity_2} (Color online) Intensities $I/I_0$ for different
absolute values of the OAM number
$L=|l|$ in the $\mathrm{\hat {e}}_{X}$-$\mathrm{\hat {e}}_{Y}$-plane. The reference intensity
is $I_0=2\cdot 10^{14}\,\mathrm{W}/\mathrm{cm}^2$.}
\ef
Varying the OAM number the spatial profile of the laser beam changes, in particular
the radial distance where the peak intensity is located is shifted.
The remaining model parameters have been chosen as follows:
The damping constant is set to $\gamma=10^{15}\,\mathrm{s}^{-1}$, the eigenfrequency
takes $\omega_0=4.650\cdot 10^{15}\,\mathrm{s}^{-1}$ (corresponding to $\hbar\,\omega_0=3.06\,\mathrm{eV}$)
and the external frequency is $\Omega=2.325\cdot 10^{15}\,\mathrm{s}^{-1}$ (corresponding to
$\hbar\,\omega_0=1.53\,\mathrm{eV}$).
The eigenfrequency represents the energy gap leading to transparency of the magnetic garnet film
for the fundamental beam \cite{Hansteen:PhysRevB:73:014421:2006}.
Furthermore, the frequencies are chosen such that $\omega_0=2\,\Omega$.
Following \cite{Boyd:NonlinearOptics:Book:2008} the anharmonic parameter can be estimated
from $a=\omega_0^2/d$, where $d$ is the lattice constant.
Here, the lattice constant is $d=1.25\,\mathrm{nm}$ as to simulate
yttrium iron garnet.
Further, right-handed circularly polarized light ($\sigma =+1$) has been considered and the number of
radial nodes and the beam waist have been set to $n=0$ and $w_0=2\,\mu\mathrm{m}$, respectively.
The quantitative values of the elements of the coupling tensor $\Gamma_{\alpha\beta\gamma}$
have been chosen based on the values for the susceptibility $\chi_{\alpha\beta\gamma}$ experimentally
determined and reported in \cite{Hansteen:PhysRevB70:094408:2004}.
Although, the exact numerical values with corresponding units could not be deduced from
\cite{Hansteen:PhysRevB70:094408:2004},
we adopted the relations between the different tensor components mentioned in this article.
In the present model the tensor components $\Sigma_\alpha$ always occur in product with the saturation
magnetization $M_s$.
We have chosen $\Sigma_1\,M_s=-0.02$, $\Sigma_2\,M_s=0.2$ and $\Sigma_3\,M_s=0.1$.
To model magnetization dynamics we made use of the Gilbert damping parameter $\alpha=10^{-4}$.

\section{Results and Discussion\label{Sec_Results}}

The investigated laser beam possesses a
specific light polarization $\sigma$ associated with the  spin degrees of freedom of the photons,
and an orbital angular momentum $l$, the effect of the latter is the prime  focus of our study.
The results presented in the following were all obtained using right-handed circularly polarized light,
i.e. $\sigma =+1$.
Important differences to the case of left-handed circularly polarized
light ($\sigma =-1$) will be pointed out.
Furthermore, in what follows we will present our results in terms of the
absolute value of the OAM-number introduced previously as $L=|l|$.

Typical trajectories of the first-order and second-order electron loops are shown in
Fig.~\ref{Fig_Loop_Different_L},
where, for the sake of clarity, we assumed an infinite pulse length for the illustrations.
However, all subsequent graphics are created taking into account the finite pulse
width $\tau_p=10\,\mathrm{fs}$.
\bef[h!]
\centering
\subfigure[$l=0$]{
\includegraphics[width=6cm]{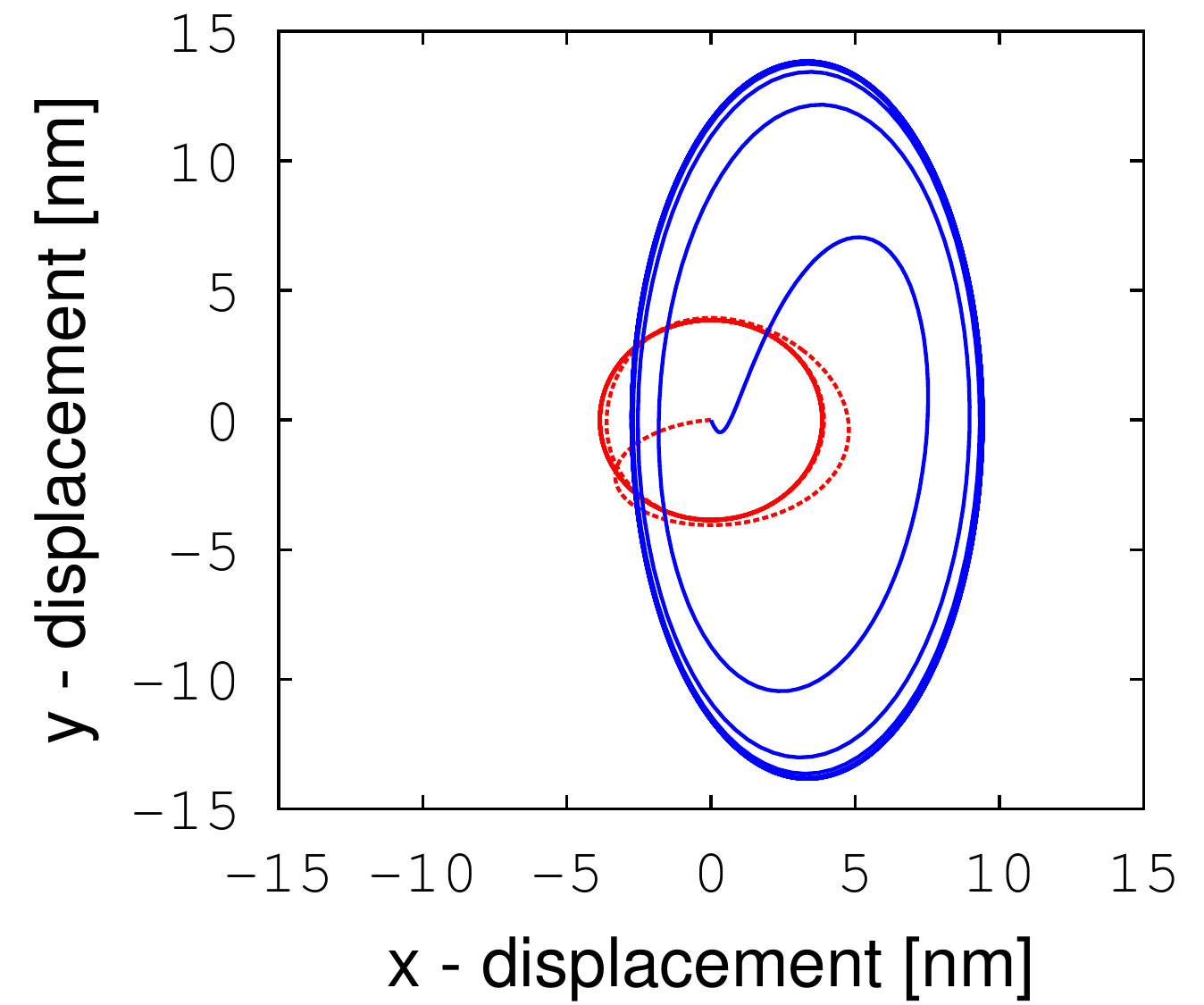}}
\subfigure[$l=4$]{
\includegraphics[width=6cm]{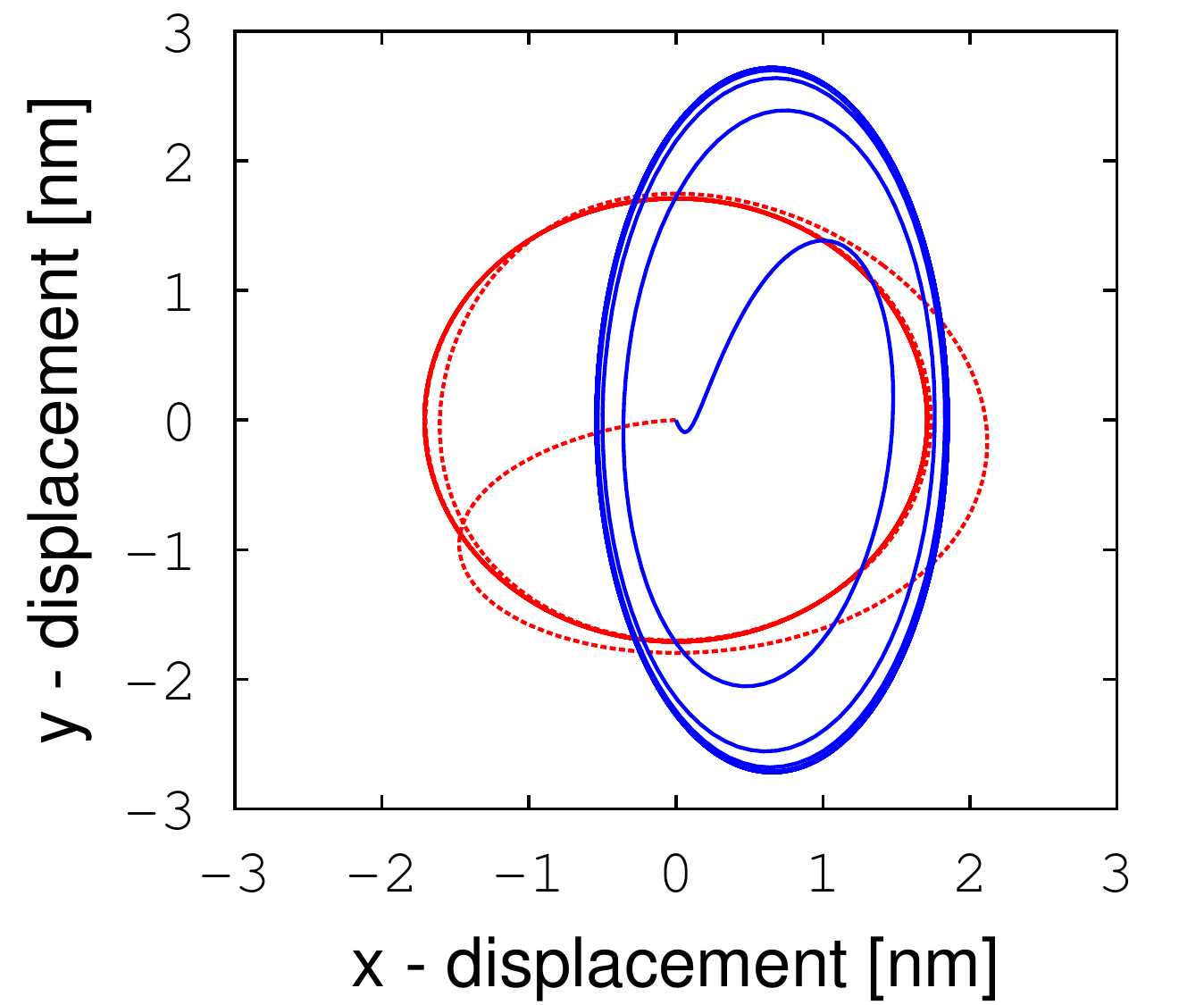}}
\subfigure[$l=6$]{
\includegraphics[width=6cm]{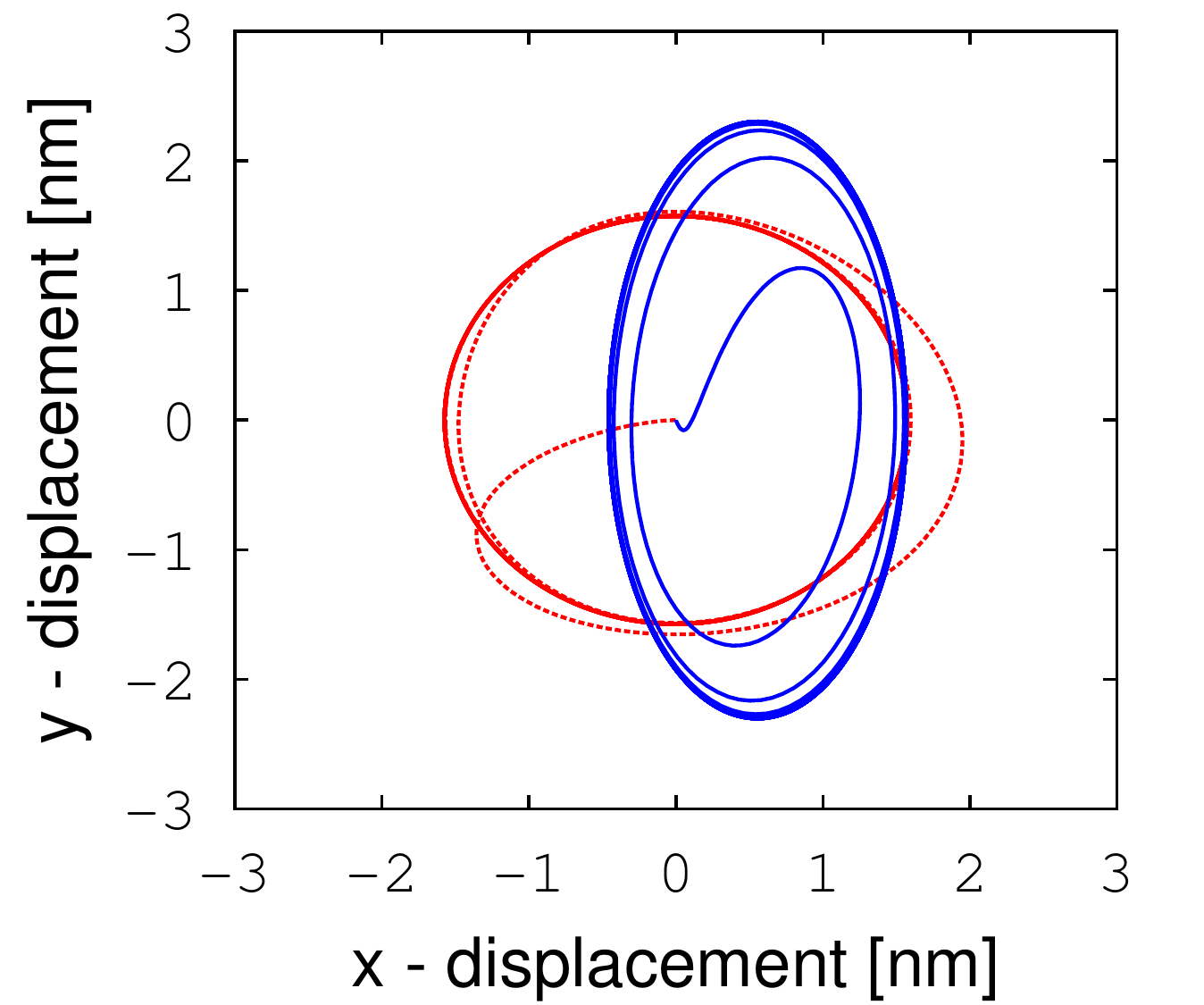}}
\caption{\label{Fig_Loop_Different_L} (Color online) Single electron current loops depending on the
OAM number $l$ for $I=2\cdot 10^{16}\,\mathrm{W}/\mathrm{cm}^2$ at $t=0$.
The first-order solution is shown as dashed (red) line and the second-order solution
is shown as solid (blue) line. Right-handed circularly polarized light ($\sigma=+1$) is considered.
Each trajectory corresponds to the electron displacement at the position of maximum laser intensity
which differs depending on $l$.}
\ef
The first-order loop (dashed red line) is described by a circular motion whereas
the second-order displacement (solid blue line) follows an ellipse.
The elliptic motion arises from the presence and the direction of the magnetization.
The direction of rotation is determined by the helicity $\sigma$
of the light pulse.
Here we assumed $\sigma =+1$.
Deviations from the circular and elliptic motions result from the transient solutions.
These solutions affect the motion of the electron significantly only during the first cycle
(orbital period $\simeq 2.7\,\mathrm{fs}$).
As a consequence of the symmetry the directions of rotation of the first-order and
second-order loops are opposite.
The first-order loop rotates anti-clockwise and the second-order motion is a clockwise rotation.
Therefore, for the case considered here (magnetic garnet films, $4mm$-symmetry)
second-order effects will lead to a weakening of the first-order
generated magnetic field
because the direction of rotation determines
the sign of the light-induced magnetic field.
This may not be the general case.
Depending on the material properties and on the experimental setup
an enhancement of the first-order generated magnetic field by second-order effects
might be observed as well.
Further, the aforementioned dc-shift of the single-electron current loop in $\mathrm{x}$-direction is
another nonlinear effect and shown in Fig.~\ref{Fig_Loop_Different_L}.
The direction of this dc-shift depends on the direction of the in-plane magnetization
but does not
alter the oscillatory motion of the electron.
Hence, the magnitude of the light-induced magnetic field will not be influenced by optical rectification.
Note that the sign of the OAM number $l$ does not influence the direction of rotation of the electron
but determines the phase $\propto \exp[\mathrm{i}\,\phi l]$ depending on the position $(r,\phi)$
at which the electron resides
in the frame of the light beam before the light pulse is applied.
The electronic displacement depicted in Fig.~\ref{Fig_Loop_Different_L} refers to a peak intensity (at $t=0$)
of $2\cdot 10^{16}\,\mathrm{W}/\mathrm{cm}^2$. For lower intensities the effect shown and discussed above
occurs as well but
the second-order contribution to the total electron displacement is much lower than
the first-order contribution.
As the first- and second-order generated magnetic fields are proportional to the square of first- and second-order
electronic displacements, respectively, the ratio $\left|B^{(2)}_{\mathrm{ind}}/B^{(1)}_{\mathrm{ind}}\right|$ should
decrease for decreasing intensities.
This effect is shown in Fig.~\ref{B_Relation_von_L} for different values of $L$.
\bef[t]
\centering \includegraphics[width=8.5cm]{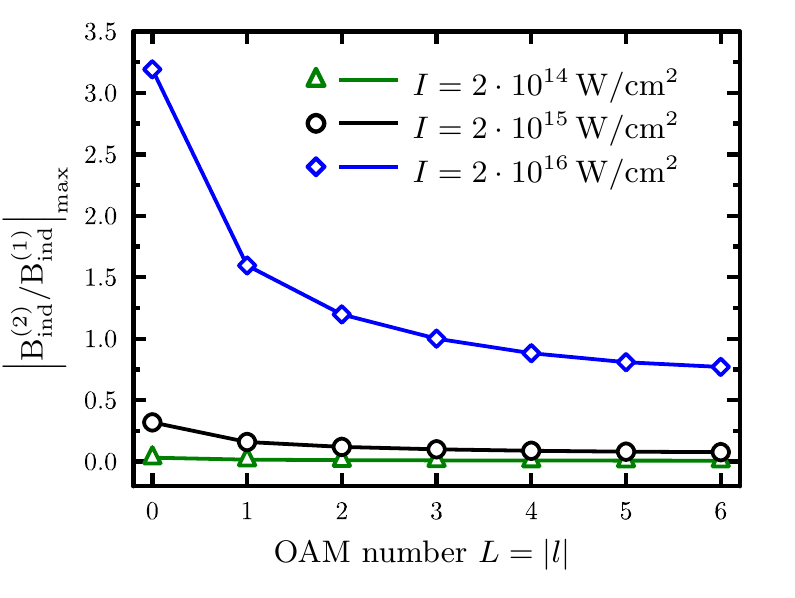}
\caption{(Color online) $L$-dependence of the ratio between second-order and first-order
light-induced magnetic fields for different intensities (calculated at $t=0$).
The index $max$ refers to the maximum value which is taken at $t=5.4\,\mathrm{fs}$ the time
at which the first value of the light-induced magnetic field is calculated (see text for further explanation).
Right-handed circularly polarized light is applied ($\sigma=+1$).}
\label{B_Relation_von_L}
\ef
Further, it can clearly be seen that the ratio shown in Fig.~\ref{B_Relation_von_L} is independent of
$L$ if the intensity goes to zero.

The total optically-generated magnetic field $B^{(1)}_{\mathrm{ind}}+B^{(2)}_{\mathrm{ind}}$ is drawn
in Fig.~\ref{B_Sum_von_t} for different intensities and $L$-values.
\bef[t]
\centering
\subfigure[$I/I_0=1$]{
\includegraphics[width=6.3cm]{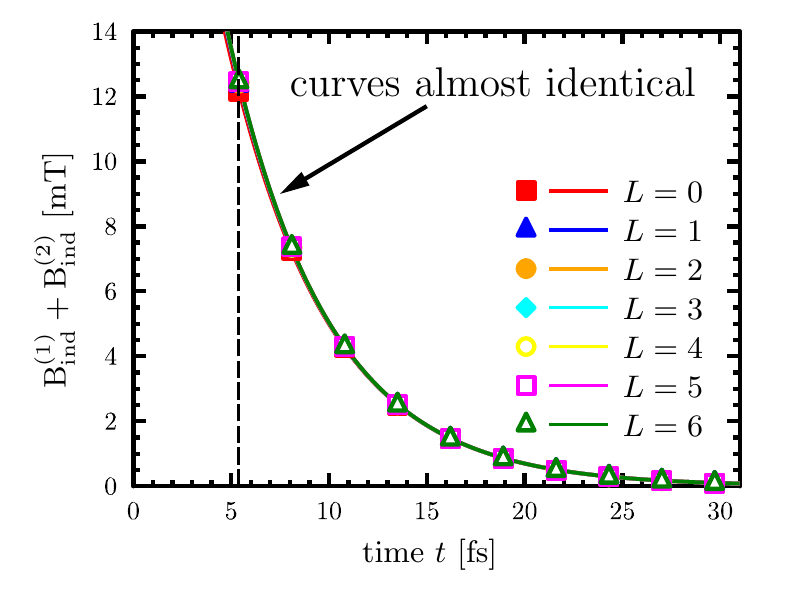}}
\subfigure[$I/I_0=10$]{
\includegraphics[width=6.3cm]{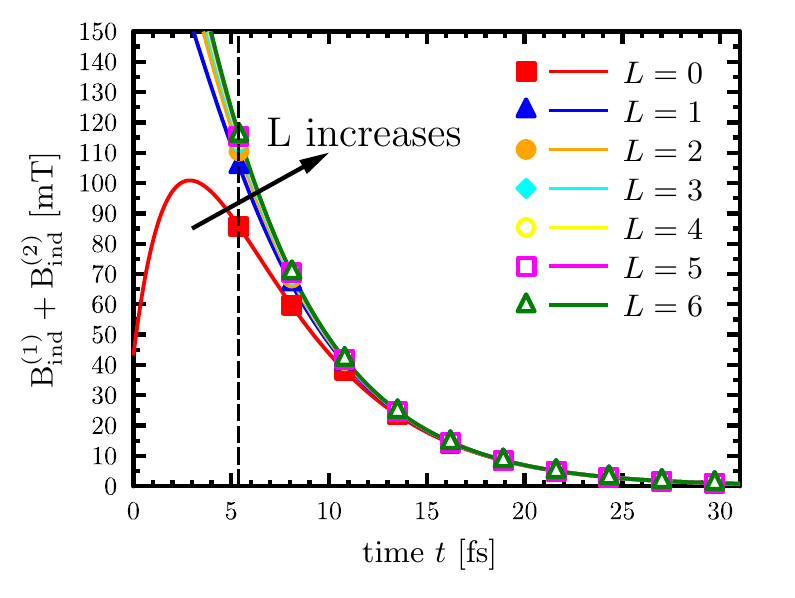}}
\subfigure[$I/I_0=100$]{
\includegraphics[width=6.3cm]{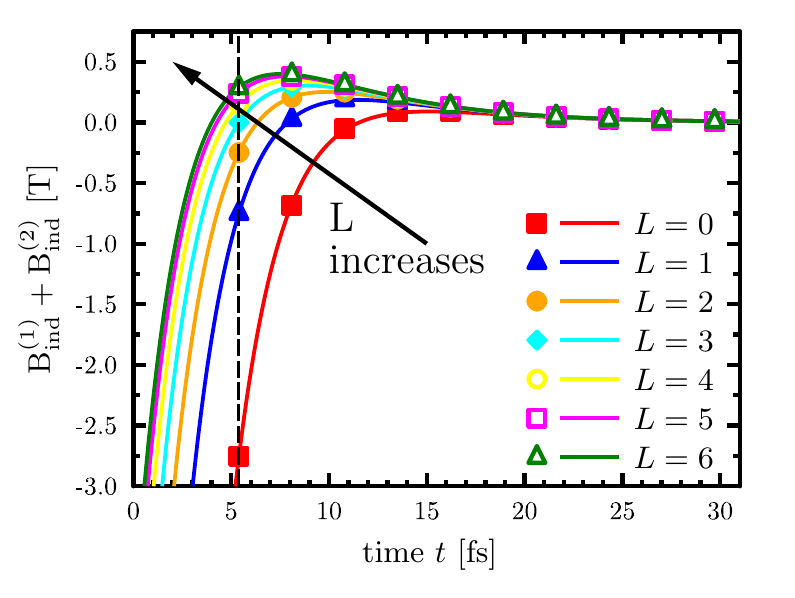}}
\caption{(Color online) Time dependence of the total light-induced magnetic field
for different intensities and absolute values of the OAM number $L=|l|$.
The dashed vertical line at $t=5.4\,\mathrm{fs}$ corresponds to two orbital periods
as described in section \ref{Sec_Opt_Gen_Field}.
The reference intensity
is $I_0=2\cdot 10^{14}\,\mathrm{W}/\mathrm{cm}^2$.
Right-handed circularly polarized light is applied ($\sigma=+1$).}
\label{B_Sum_von_t}
\ef
The markers represent the data points calculated after each full cycle the electrons finalize and
the solid lines correspond to a fit of the superposition of two exponential functions
(referring to $B^{(1)}_{\mathrm{ind}}$ and $B^{(2)}_{\mathrm{ind}}$) to the data points.
Obviously, the values of the total magnetic field are always positive in Figs.~\ref{B_Sum_von_t}\,(a)
and \ref{B_Sum_von_t}\,(b). Whereas, at sufficient high intensities the total magnetic field
can undergo a sign change as shown in Fig.~\ref{B_Sum_von_t}\,(c).
Also note the different orders of magnitude in Figs.~\ref{B_Sum_von_t}\,(a) - \ref{B_Sum_von_t}\,(c).
Again it can be observed that the influence of $L$ on the shape of the total magnetic field enhances
as the intensity increases.
For a quantitative comparison we refer to table~\ref{table1}.
\begin{table}
\caption{\label{table1}Listing of the maximum values of the superposition of first- and second-order
optically-generated magnetic fields. These values are calculated at $5.4\,\mathrm{fs}$ after
the application of the laser pulse. Values are given in $10^{-3}\,\mathrm{Tesla}$.
The value of the reference intensity is $I_0=2\cdot 10^{14}\,\mathrm{W}/\mathrm{cm}^2$. $L=|l|$ is
the absolute value of the OAM number.}
\begin{indented}
\item[]\begin{tabular}{@{}l|c c c c c c c }
\br
 & $L=0$ & $L=1$ & $L=2$ & $L=3$ & $L=4$ & $L=5$ & $L=6$\\ \hline
$I/I_0=1$ & $12$ & $12$ & $12$ & $12$ & $12$ & $12$ & $12$ \\
$I/I_0=10$ & $86$ & $106$ & $111$ & $113$ & $116$ & $115$ & $116$ \\
$I/I_0=100$ & $-2752$ & $-748$ & $-248$ & $-1$ & $148$ & $240$ & $289$ \\
\br
\end{tabular}
\end{indented}
\end{table}
In addition, from a fitting of the curves  in Fig.~\ref{B_Sum_von_t} we deduce
the decay time $\tau^{(1)}=5.05\,\mathrm{fs}$ referring to $B^{(1)}_{\mathrm{ind}}(t)$
and the decay time $\tau^{(2)}=2.59\,\mathrm{fs}$ referring to $B^{(2)}_{\mathrm{ind}}(t)$.
Taking into account that the exponential decay time of the external field pulse $E(t)$ is assumed to
be $\tau_p=10\,\mathrm{fs}$
and that the first-order magnetic field $B^{(1)}_{\mathrm{ind}}(t)\propto E^2(t)$
and the second-order magnetic field $B^{(2)}_{\mathrm{ind}}(t)\propto E^4(t)$,
these values are close to $\tau_p/2$ and $\tau_p/4$ which can be expected theoretically.

To further underline that a change in the optically-generated magnetic field induced by
varying the OAM in terms of the number $L$ is clearly a second-order effect, we present Fig.~\ref{B_1_B_2_von_t}
where the time dependencies of the optically-generated first-order and second-order magnetic fields,
$B^{(1)}_{\mathrm{ind}}$ and $B^{(2)}_{\mathrm{ind}}$, are depicted for different $L$.
\bef[b]
\centering \includegraphics[width=8.5cm]{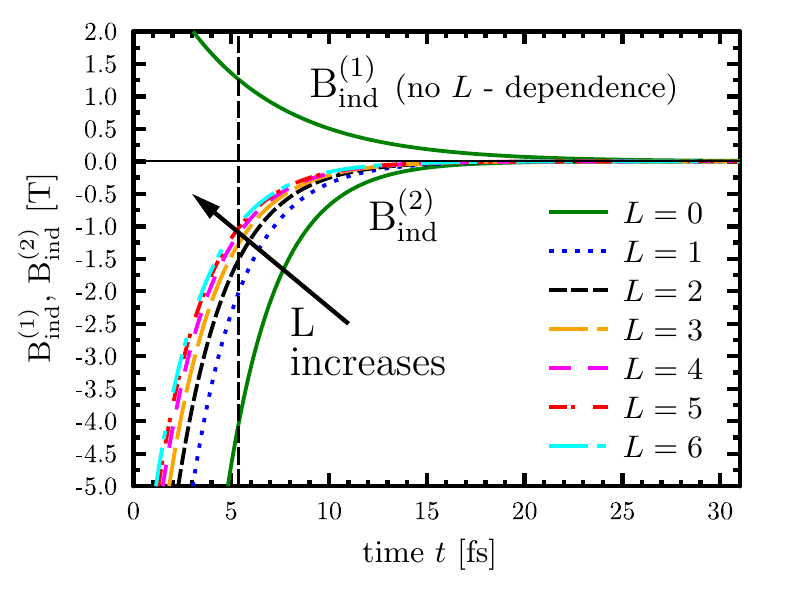}
\caption{(Color online) Comparison of the time dependence of first- and  second-order light-induced
magnetic fields
for different absolute values of the OAM number $L=|l|$ and $I=2\cdot10^{16}\,\mathrm{W}/\mathrm{cm}^2$.
The dashed vertical line at $t=5.4\,\mathrm{fs}$ corresponds to two orbital periods
as described in section \ref{Sec_Opt_Gen_Field}.
Right-handed circularly polarized light is applied ($\sigma=+1$).}
\label{B_1_B_2_von_t}
\ef
As is visible $B^{(1)}_{\mathrm{ind}}(t)$ does not depend on $L=|l|$ whereas the magnitude and the sign of
$B^{(2)}_{\mathrm{ind}}(t)$ are significantly influenced by changing $L$.

Additionally, we point out that in the event  the laser beam is left-handed circularly polarized ($\sigma=-1$)
both the first-order and the second-order
generated magnetic fields will change sign.
Consequently, the total optically-induced magnetic field will also undergo a sign change (not shown).

In the present model the second harmonics generation is solely due to the presence of
a magnetization.
Vice versa the second harmonics will produce the light-induced magnetic field $B^{(2)}_{\mathrm{ind}}$
which influences the magnetic structure.
To investigate the effect of the total light-induced magnetic field on the magnetic system
we solved the Landau-Lifshitz-Gilbert in Eq.~(\ref{LLGeq}) numerically taking into account the effective magnetic
field in Eqs.~(\ref{MagField}) and (\ref{Beff}).
A static magnetic field $\mathbf{B}_0=B_0\,\mathbf{\hat {e}}_{Y}$ with $B_0=250\,\mathrm{mT}$ was applied.
Whereas, the light-induced magnetic fields point into the $\pm\mathrm{Z}$-direction.
The corresponding results are shown in Fig.~\ref{Fig_LLG_1} for different intensities.
\bef[b]
\centering \includegraphics[width=8.5cm]{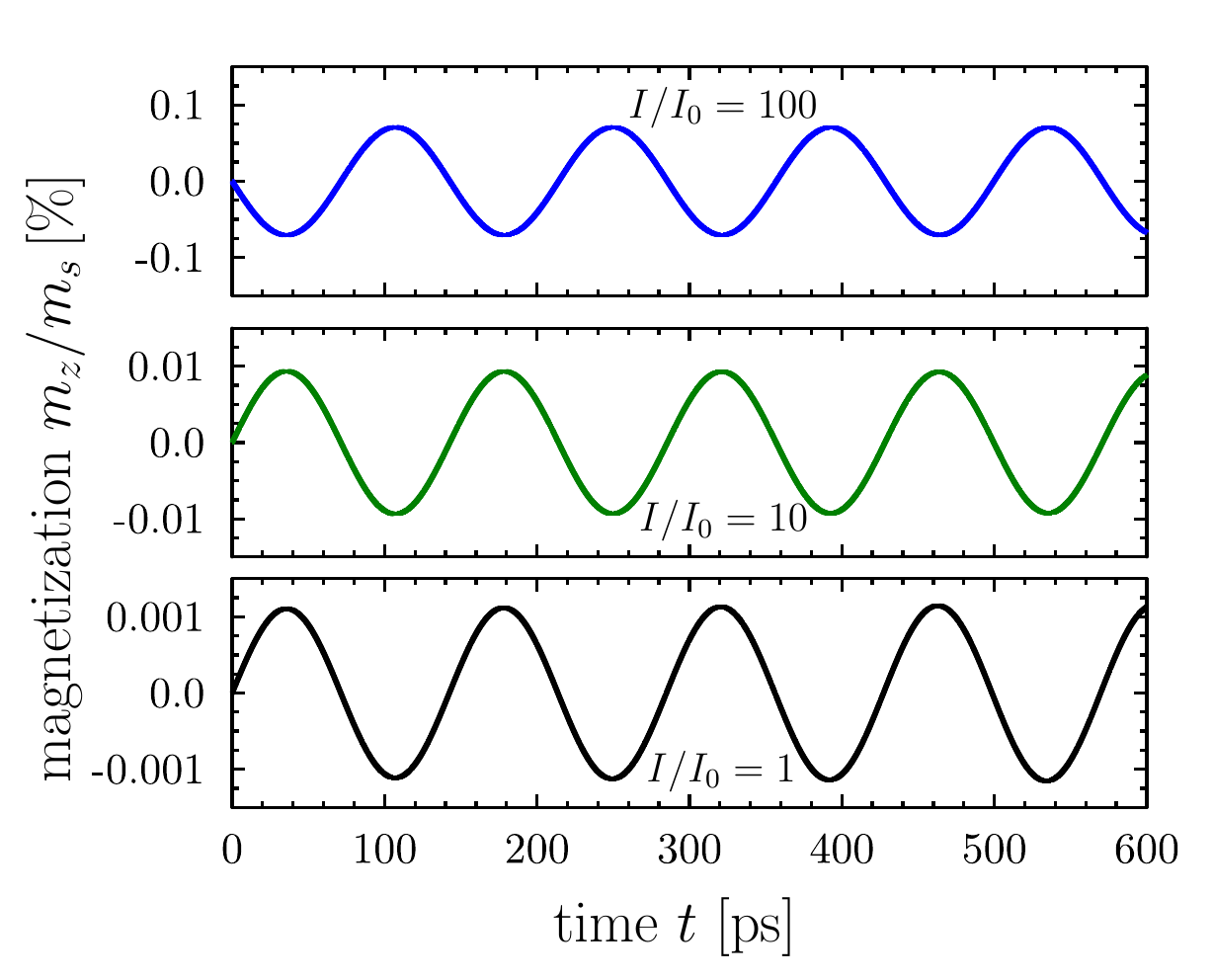}
\caption{Numerical solution of the Landau-Lifshitz-Gilbert equation for different intensities
and $L=0$.
The reference intensity
$I_0=2\cdot 10^{14}\,\mathrm{W}/\mathrm{cm}^2$ as well as $I=I(E_0,\,t=0)$ which is varied by varying
the electrical field strength $E_0$ are taken at $t=0$.
The applied effective magnetic field is given in
Eqs.~(\ref{MagField}) and (\ref{Beff}).}
\label{Fig_LLG_1}
\ef
Obviously, the light pulses induce magnetization dynamics in terms of small-angle magnetic excitations.
The precession amplitude can be varied by changing the intensity of the laser pulse.
Regarding this, a variation of two orders of magnitude of the precession amplitudes is
observed for $1\leq I/I_0 \leq 100$, with $2\cdot 10^{14}\,\mathrm{W}/\mathrm{cm}^2$.
For $I/I_0=100$ the total light-induced magnetic field is negative which is indicated by the excitation of the
out-of-plane component of the magnetization $m_z$.
As can be seen in Fig.~\ref{Fig_LLG_1} the oscillations of $m_z$ start into the opposite direction
compared with the cases $I/I_0=1$ and $I/I_0=10$.

Viewing  the theoretical results from an experimental side
\cite{Hansteen:PhysRevLett:95:047402:2005,Hansteen:PhysRevB:73:014421:2006},
the observed precession amplitudes are not in the experimentally determined
order of magnitude.
On the one hand, this discrepancy originates from the pulse width $\tau_p=10\,\mathrm{fs}$ of the single pulse,
as applied in the present model.
Because of the different time scales of the optical pulse and the precessional dynamics of
the magnetization, the short presence of the light-induced magnetic field
leads to small amplitudes of the precessing magnetization in the numerical modeling.
For pulses which live longer, e.g. considering a pulse width $\tau_p=100\,\mathrm{fs}$
or a sequence of pulses the precession amplitude would increase.
On the other hand, due to computational reasons we assumed a beam waist of $w_0=2\,\mu\mathrm{m}$.
Experimentally a typical value for the beam waist is approximately $100-200\,\mu\mathrm{m}$.
However, as shown theoretically \cite{Thakur_OptExpr_18_27691_2010}, it is  possible to exploit self-focussing effects to decrease the waist (and increase the intensity) quite significantly.
A larger illuminated area may increase the magnitude of the magnetic field generated by
the laser light as well.
\bef[t]
\centering \includegraphics[width=8.5cm]{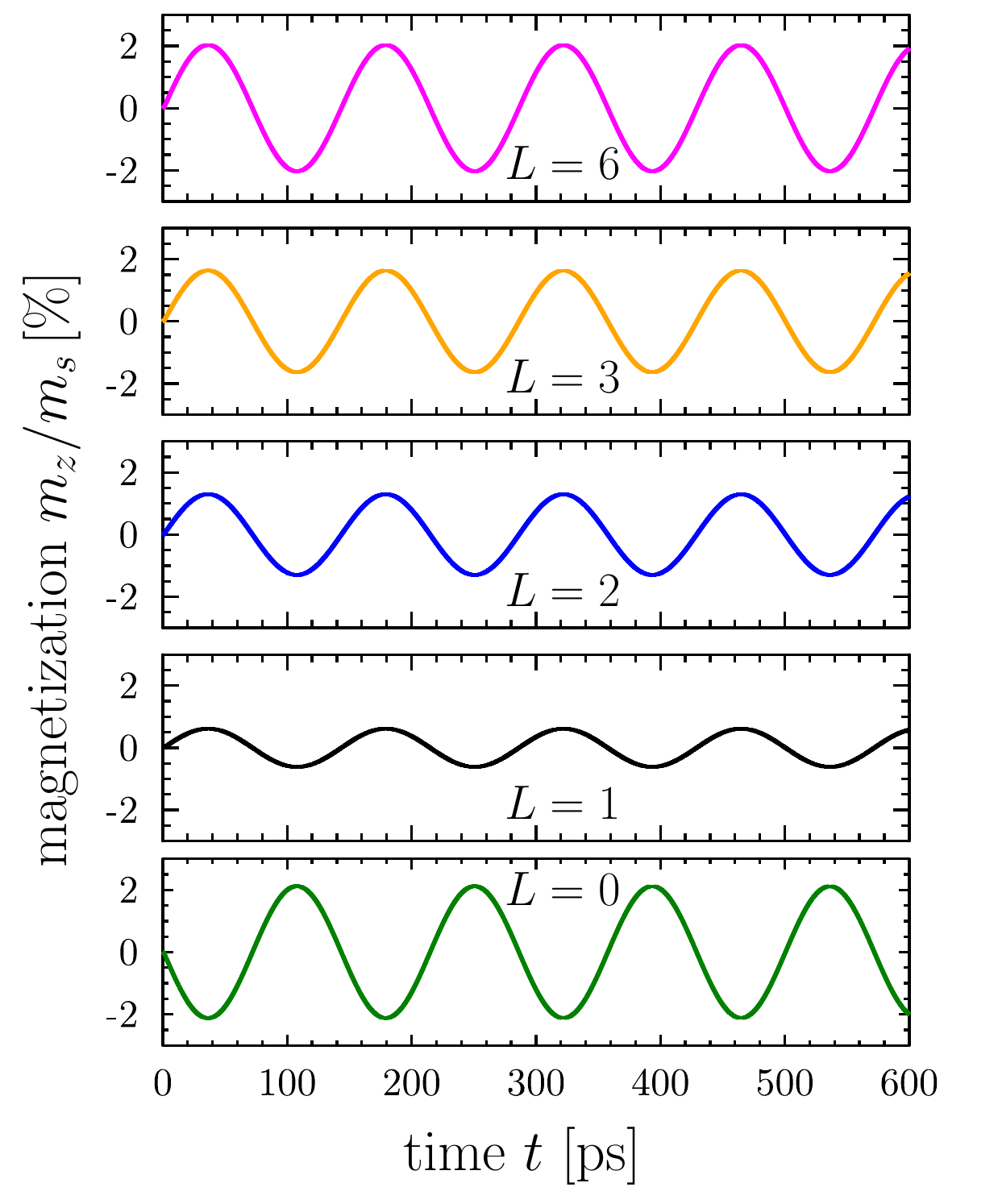}
\caption{Numerical solution of the Landau-Lifshitz-Gilbert equation for a sequence of $30$ pulses with
peak intensity $I=2\cdot10^{16}\,\mathrm{W}/\mathrm{cm}^2$ at $t=0$.}
\label{Fig_LLG_2}
\ef
Fig.~\ref{Fig_LLG_2} shows the predictions for the
magnetization dynamics when applying a sequence of $30$ identical pulses with
pulse width $\tau_p=10\,\mathrm{fs}$ and a time delay between two subsequent pulses of
$\Delta t=50\,\mathrm{fs}$ assuming a peak intensity of $I=2\cdot10^{16}\,\mathrm{W}/\mathrm{cm}^2$
at $t=0$.
The $L=0$ subfigure in Fig.~\ref{Fig_LLG_2} should be compared with the
$I/I_0=100$ subfigure in Fig.~\ref{Fig_LLG_1}.
As can be seen, the changes described above lead to a gain of one order of magnitude with
regard to the precessional amplitude.
It is moreover obvious that the parameter $L$ expressing the OAM of the laser beam
determines the sign of the light-induced magnetic field as well as its magnitude.
The sign change is indicated by the excitation of the magnetization to the opposite direction
in the $L=0$ case compared to the $L>0$ cases.
We will discuss below that this is a more subtle issue than obvious at first glance.
Additionally, going from $L=0$ to $L=1$ the absolute value of the magnetization amplitude decreases
and increases when further increasing $L>1$.
This behavior can be explained by looking at the curves plotted in Fig.~\ref{B_Relation_von_L} and
and the values given in table~\ref{table1}.
Considering the $L=0$ case a large negative magnetic field is observed after the first full cycle
of the electron motion at $t=5.4\,\mathrm{fs}$, see table~\ref{table1}, followed by two smaller values
after the second loop (at $t=8.1\,\mathrm{fs}$) and the third loop (at $t=10.8\,\mathrm{fs}$).
All subsequent values calculated at increments of $\Delta t=2.7\,\mathrm{fs}$ are positive but smaller in magnitude.
However, the total magnetic field exhibits negative values for $L=1,2,3$ as well, but, in contrast to the $L=0$
behavior, this cannot be monitored by observing the magnetization dynamics on the picosecond-timescale as
shown in the corresponding simulations in Fig.~\ref{Fig_LLG_2}.
For the $L=1,2,3$ cases the optically-induced magnetic field becomes positive for all subsequent
cyclic motions of the electrons.
This circumstance cancels the negative field corresponding to $t=5.4\,\mathrm{fs}$ as the second-order
magnetic field decays twice as fast as the first-order magnetic field, i.e. $\tau^{(2)}\approx \tau^{(1)}/2$.
The occurrence that the $L=1$ case offers the smallest amplitude of the magnetization of
all shown examples in Fig.~\ref{Fig_LLG_2} arises from those
cancelation effects during the first few time steps.
For a comparison of the first- and second-order magnetic fields at $t=5.4\,\mathrm{fs}$ for different $L$
we again refer to Fig.~\ref{B_Relation_von_L} (cf. also the behavior of the total magnetic field shown
in Fig.~\ref{B_Sum_von_t} (c)).

We infer from these simulations that the light
beams with OAM may in principle generate fs magnetic field pulses that can be used to
coherently control the magnetization in the manner suggested in
Ref.\cite{Sukhov_PRL_102_057204_2009,Sukhov_PRB_79_134433_2009,Sukhov_AppPhysA_98_837_2010}, where it
was also shown that on this time scale the control scheme is robust to thermal fluctuations.

Finally, we briefly address the influence of the in-plane orientation of the magnetization.
Rotating the magnetization by an angle $\phi_\mathrm{rot}$ away from the
$\mathrm{\hat {e}}_{Y} = \mathrm{y}$-direction would
change the equation of motion of the second-order displacement in Eq.~(\ref{EOM_2ndOrder})
because of the contribution of the tensor components $\Gamma_{\alpha\beta\gamma x}$
in Eqs.~(\ref{NonzeroComponents1})
and (\ref{EOM_2ndOrder})
which would vanish for $\phi_\mathrm{rot}=0$, that is $\mathbf{M} = M_s\,\mathbf{\hat {e}}_{Y} = M_s\,\mathbf{y}$
and $M_\mathrm{x}=0$.
For a finite angle of rotation the magnetization components $M_\mathrm{x}=M_s\,\cos\,\phi_\mathrm{rot}$
and $M_\mathrm{y}=M_s\,\sin\,\phi_\mathrm{rot}$ should be considered instead.
This would result in a dc-shift of the single-electron current loop either in $\mathrm{x}$-direction or in $\mathrm{y}$-direction
or in both, depending on the rotation angle. In contrast, if $\phi_\mathrm{rot}=0$ as applied in the present study, a dc-shift
is only observed in $\mathrm{x}$-direction.
Although the detailed dependence of the second-order displacement and the magnetic excitations on the rotation angle $\phi_\mathrm{rot}$
is an interesting aspect, this is beyond the scope of the present paper.

\section{Conclusion\label{Sec_Conclusion}}

Nonthermal opto-magnetic effects are intriguing phenomena in condensed matter physics that, on the one hand,
bear high potential
for promising magnetic devices and applications and pose interesting questions
for theory.
In the present study we worked out the role of nonlinear effects which are not negligible for intensities
in the range of $2\cdot 10^{14}-2\cdot 10^{16}\,\mathrm{W}/\mathrm{cm}^2$.
Utilizing a classical treatment of the laser-driven carrier in a given symmetry environment,
it is possible to separate the electron motion into a first-order
displacement that is directly proportional to the electrical field of the optical pulse and
a second-order displacement that depends on the square of the electrical field.
Considering magnetic insulators with a certain symmetry configuration we found that both the first-order
and the second-order electron displacement create current loops that generate the light-induced
magnetic fields $B^{(1)}_{\mathrm{ind}}$ and $B^{(2)}_{\mathrm{ind}}$.
As the direction of rotation of the first-order electron displacement always
opposes that of the second-order displacement, the light-induced fields $B^{(1)}_{\mathrm{ind}}$
and $B^{(2)}_{\mathrm{ind}}$ also carry opposite signs.
Applying Laguerre-Gaussian laser beams which carry orbital angular momentum characterized
by the number $l$, we showed that the optically-generated magnetic
field can be controlled by a variation of $L=|l|$.
Thus, in the classical model presented in this paper the light-induced magnetic field
depends on the absolute value of the OAM number but is independent of its sign.
Hence, the sign of the total magnetic field can be positive or negative and the strength of the
light-induced magnetic field can reach several Tesla.
In the present study we investigated nonlinear effects for $0\leq L\leq 6$.
Increasing the illuminated area, larger values of $L$ can be studied as well.

Finally, we note that the here reported effect based on the generation of first-order and second-order
magnetic fields with opposite signs may change for other crystallographic and/or magnetization
induced symmetries.
We assume that an enhancement of the first-order magnetic field by the second-order field may
occur also in systems with different symmetries.
However, nonlinear opto-magnetic effects as investigated in the present paper still need to be
verified experimentally.\\
To verify the predicted phenomena we suggest an optical pump-probe experiment similar to that described in \cite{Krilyuk:RevModPhys:p2731:2010}
for studying the magnetization dynamics.
A high-intensity circularly polarized laser pulse that carries OAM could be used
to excite magnetization dynamics which could then be measured by probing
the motion of the magnetization vector with linearly polarized light pulses of much lower intensity.
From determining the excitation angle the light-induced magnetic field could be derived.
In particular, the variation of the intensity and OAM of the pump pulse
should lead to a point where
first-order and second-order light induced fields, which have opposite signs, are almost equal in magnitude.
If this compensation point is reached the amplitude of the magnetic excitations should be
reduced significantly, compare Fig.~\ref{Fig_LLG_2}.
Therefore, by probing the magnetic changes those intensities and OAM-numbers could be derived.

We hope our investigations will stimulate further studies related to optical as well
as electron vortex beams.
In particular, it was shown recently within a quantum mechanical theory
\cite{Lloyd_PRL_108_074802_2012}
that electron vortex beams are able to exchange orbital angular momentum with the electrons of the irradiated
atom in the dipole as well as quadrupole transitions.
Further, the transitions were shown to be independent of the sign of the OAM number which coincides with
our results for optical vortex beams interacting with magnetic matter. This similarity is not surprising as in the
optical limit charge particle scattering at small momentum transfer amounts to the interaction with photons.

\section{Acknowledgments}

We benefited from valuable discussions with S Trimper, L Chotorlishvili and R W Chantrell.

\section*{References}

\bibliography{References_NIF_2}

\begin{thebibliography}{10}

\bibitem{MFundNano:2006}
J.~St\"ohr and H.~C. Siegmann.
\newblock {\em Magnetism - From Fundamentals to Nanoscale Dynamics}.
\newblock Springer, Berlin, 2006.

\bibitem{Krilyuk:RevModPhys:p2731:2010}
A.~Kirilyuk, A.V. Kimel, and T.~Rasing.
\newblock Ultrafast optical manipulation of magnetic order.
\newblock {\em Rev. Mod. Phys.}, 82:2731--2784, 2010.

\bibitem{Beaurepaire:PhysRevLett76:4250:1996}
E.~Beaurepaire, J.-C. Merle, A.~Daunois, and J.-Y. Bigot.
\newblock Ultrafast spin dynamics in ferromagnetic nickel.
\newblock {\em Phys. Rev. Lett.}, 76:4250, 1996.

\bibitem{Kazantseva:EPL81:27004:2008}
N.~Kazantseva, U.~Nowak, R.~W. Chantrell, J.~Hohlfeld, and A.~Rebei.
\newblock Slow recovery of the magnetisation after a sub-picosecond heat pulse.
\newblock {\em EPL}, 81:27004, 2008.

\bibitem{Koopmans:PhysRevLett:85:844:2000}
B.~Koopmans, M.~van Kampen, J.~T. Kohlhepp, and W.~J.~M. de~Jonge.
\newblock Ultrafast magneto-optics in nickel: Magnetism or optics?
\newblock {\em Phys. Rev. Lett.}, 85:844--847, 2000.

\bibitem{Vahaplar:PhysRevLett:103:117201:2009}
K.~Vahaplar, A.~M. Kalashnikova, A.~V. Kimel, D.~Hinzke, U.~Nowak,
  R.~Chantrell, A.~Tsukamoto, A.~Itoh, A.~Kirilyuk, and Th. Rasing.
\newblock Ultrafast path for optical magnetization reversal via a strongly
  nonequilibrium state.
\newblock {\em Phys. Rev. Lett.}, 103:117201, 2009.

\bibitem{Radu:Nature472:205:2011}
I.~Radu, K.~Vahaplar, C.~Stamm, T.~Kachel, N.~Pontius, H.~A. Durr, T.~A.
  Ostler, J.~Barker, R.~F.~L. Evans, R.~W. Chantrell, A.~Tsukamoto, A.~Itoh,
  A.~Kirilyuk, Th. Rasing, and A.~V. Kimel.
\newblock Transient ferromagnetic-like state mediating ultrafast reversal of
  antiferromagnetically coupled spins.
\newblock {\em Nature}, 472:205, 2011.

\bibitem{Ostler:NatComm:3:666:2012}
T.A. Ostler, J.~Barker, R.F.L. Evans, R.W. Chantrell, U.~Atxitia,
  O.~Chubykalo-Fesenko, S.~El~Moussaoui, L.~Le~Guyader, E.~Mengotti, L.J.
  Heyderman, F.~Nolting, A.~Tsukamoto, A.~Itoh, D.~Afanasiev, B.A. Ivanov, A.M.
  Kalashnikova, K.~Vahaplar, J.~Mentink, A.~Kirilyuk, Th. Rasing, and A.V.
  Kimel.
\newblock Ultrafast heating as a sufficient stimulus for magnetization reversal
  in a ferrimagnet.
\newblock {\em Nat. Commun.}, 3:666, 2012.

\bibitem{Vahaplar:PhysRevB:85:104402:2012}
K.~Vahaplar, A.~M. Kalashnikova, A.~V. Kimel, S.~Gerlach, D.~Hinzke, U.~Nowak,
  R.~Chantrell, A.~Tsukamoto, A.~Itoh, A.~Kirilyuk, and Th. Rasing.
\newblock All-optical magnetization reversal by circularly polarized laser
  pulses: Experiment and multiscale modeling.
\newblock {\em Phys. Rev. B}, 85:104402, 2012.

\bibitem{Shen:NonlinearOptics:Book:1984}
Y.~R. Shen.
\newblock {\em The Principles of Nonlinear Optics}.
\newblock John Wiley \& Sons, 1984.

\bibitem{Kalshnikova:PhysRevLett:99:167205:2007}
A.~M. Kalashnikova, A.~V. Kimel, R.~V. Pisarev, V.~N. Gridnev, A.~Kirilyuk, and
  Th. Rasing.
\newblock Impulsive generation of coherent magnons by linearly polarized light
  in the easy-plane antiferromagnet {FeBO}$_{3}$.
\newblock {\em Phys. Rev. Lett.}, 99:167205, 2007.

\bibitem{Gridnev:PhysRevB77:094426:2008}
V.~N. Gridnev.
\newblock Phenomenological theory for coherent magnon generation through
  impulsive stimulated {R}aman scattering.
\newblock {\em Phys. Rev. B}, 77:094426, 2008.

\bibitem{Pitaeevskii:SovPhysJETP:12:1008:1961}
L.~P. Pitaeevskii.
\newblock Electric forces in a transparent dispersive medium.
\newblock {\em Sov. Phys. JETP}, 12:1008--1013, 1961.

\bibitem{Ziel:PhysRevLett:15:190:1965}
J.~P. van~der Ziel, P.~S. Pershan, and L.~D. Malmstrom.
\newblock Optically-induced magnetization resulting from the inverse faraday
  effect.
\newblock {\em Phys. Rev. Lett.}, 15:190--193, 1965.

\bibitem{Pershan:PhysRev:143:574:1966}
P.~S. Pershan, J.~P. van~der Ziel, and L.~D. Malmstrom.
\newblock Theoretical discussion of the inverse {F}araday effect, {R}aman
  scattering, and related phenomena.
\newblock {\em Phys. Rev.}, 143:574--583, 1966.

\bibitem{Landau:ElecContMed:Book:1989}
L.~D. Landau, E.M. Lifshitz, and L.P. Pitaevskii.
\newblock {\em Electrodynamics of continuous media}.
\newblock Pergamon Press, Oxford, 1989.

\bibitem{Kimel:Nature:435:2005}
A.~V. Kimel, A.~Kirilyuk, P.~A. Usachev, R.~V. Pisarev, A.~M. Balbashov, and
  Th. Rasing.
\newblock Ultrafast non-thermal control of magnetization by instantaneous
  photomagnetic pulses.
\newblock {\em Nature}, 435(7042):655--657, June 2005.

\bibitem{Hansteen:PhysRevB:73:014421:2006}
Fredrik Hansteen, Alexey Kimel, Andrei Kirilyuk, and Theo Rasing.
\newblock Nonthermal ultrafast optical control of the magnetization in garnet
  films.
\newblock {\em Phys. Rev. B}, 73:014421, 2006.

\bibitem{Steil:PhysRevB:84:224408:2011}
Daniel Steil, Sabine Alebrand, Alexander Hassdenteufel, Mirko Cinchetti, and
  Martin Aeschlimann.
\newblock All-optical magnetization recording by tailoring optical excitation
  parameters.
\newblock {\em Phys. Rev. B}, 84:224408, 2011.

\bibitem{Makino:PhysRevB:86:064403:2012}
T.~Makino, F.~Liu, T.~Yamasaki, Y.~Kozuka, K.~Ueno, A.~Tsukazaki, T.~Fukumura,
  Y.~Kong, and M.~Kawasaki.
\newblock Ultrafast optical control of magnetization in euo thin films.
\newblock {\em Phys. Rev. B}, 86:064403, 2012.

\bibitem{Mikhaylovskiy:PhysRevB:86:100405:100405}
R.~V. Mikhaylovskiy, E.~Hendry, and V.~V. Kruglyak.
\newblock Ultrafast inverse faraday effect in a paramagnetic terbium gallium
  garnet crystal.
\newblock {\em Phys. Rev. B}, 86:100405, 2012.

\bibitem{Bakunov:PhysRevB:86:134405:2012}
M.~I. Bakunov, R.~V. Mikhaylovskiy, and S.~B. Bodrov.
\newblock Probing ultrafast optomagnetism by terahertz cherenkov radiation.
\newblock {\em Phys. Rev. B}, 86:134405, 2012.

\bibitem{Hertel:JMagMagMat:303:2006:L1}
R.~Hertel.
\newblock Theory of the inverse {F}araday effect in metals.
\newblock {\em J. Mag. Mag. Mat.}, 303:L1 -- L4, 2006.

\bibitem{Zhang:JMagMagMat:321:L73:2009}
Hui-Liang Zhang, Yan-Zhong Wang, and Xiang-Jun Chen.
\newblock A simple explanation for the inverse {F}araday effect in metals.
\newblock {\em J. Mag. Mag. Mat.}, 321(24):L73 -- L74, 2009.

\bibitem{Woodford:PhysRevB:79:212412:2009}
S.~R. Woodford.
\newblock Conservation of angular momentum and the inverse {F}araday effect.
\newblock {\em Phys. Rev. B}, 79:212412, 2009.

\bibitem{Yoshino:JMagMagMat:323:2531:2011}
T.~Yoshino.
\newblock Simple theory of the inverse {F}araday effect with relationship to
  optical constants n and k.
\newblock {\em J. Mag. Mag. Mat.}, 323:2531--2532, 2011.

\bibitem{Taguchi:PhysRevB:84:174433:2011}
Katsuhisa Taguchi and Gen Tatara.
\newblock Theory of inverse {F}araday effect in a disordered metal in the
  terahertz regime.
\newblock {\em Phys. Rev. B}, 84:174433, 2011.

\bibitem{Popova:PhysRevB:85:094419:2012}
D.~Popova, A.~Bringer, and S.~Bl\"ugel.
\newblock Theoretical investigation of the inverse {F}araday effect via a
  stimulated {R}aman scattering process.
\newblock {\em Phys. Rev. B}, 85:094419, 2012.

\bibitem{Taguchi:PhysRevLett:109:127204:2012}
Katsuhisa Taguchi, Jun-ichiro Ohe, and Gen Tatara.
\newblock Ultrafast magnetic vortex core switching driven by the topological
  inverse faraday effect.
\newblock {\em Phys. Rev. Lett.}, 109:127204, Sep 2012.

\bibitem{Zhu_PRB_82_235304_2010}
Z.-G. Zhu, C.-L. Jia, and J.~Berakdar.
\newblock Proposal for fast optical control of spin dynamics in a quantum wire.
\newblock {\em Phys. Rev. B}, 82:235304, 2010.

\bibitem{Matos_PRL94_166801_2005}
A.~Matos-Abiague and J.~Berakdar.
\newblock Photoinduced charge currents in mesoscopic rings.
\newblock {\em Phys. Rev. Lett.}, 94:166801, 2005.

\bibitem{Matos_EPL_69_277_2005}
A.~Matos-Abiague and J.~Berakdar.
\newblock Ultrafast build-up of polarization in mesoscopic rings.
\newblock {\em Europhys. Lett.}, 69(2):277, 2005.

\bibitem{Moskalenko_PRB_80_193407_2009}
A.~S. Moskalenko and J.~Berakdar.
\newblock Light-induced valley currents and magnetization in graphene rings.
\newblock {\em Phys. Rev. B}, 80:193407, 2009.

\bibitem{Moskalenko_PRB_74_161303_2006}
A.~S. Moskalenko, A.~Matos-Abiague, and J.~Berakdar.
\newblock Revivals, collapses, and magnetic-pulse generation in quantum rings.
\newblock {\em Phys. Rev. B}, 74:161303, 2006.

\bibitem{Allen:PRA:45:8185:1992}
L.~Allen, M.~W. Beijersbergen, R.~J.~C. Spreeuw, and J.~P. Woerdman.
\newblock Orbital angular momentum of light and the transformation of
  laguerre-gaussian laser modes.
\newblock {\em Phys. Rev. A}, 45:8185--8189, 1992.

\bibitem{Allen:ProgOpt:OAM:291:1999}
L.~Allen, M.J. Padgett, and M.~Babiker.
\newblock The orbital angular momentum of light.
\newblock {\em Prog. Opt.}, 39:291 -- 372, 1999.

\bibitem{Molina:NatPhys:3:305:2007}
Gabriel Molina-Terriza, Juan~P. Torres, and Lluis Torner.
\newblock Twisted photons.
\newblock {\em Nat. Phys.}, 3(5):305--310, 2007.

\bibitem{Bliokh_PhysRevLett_99_190404_2007}
Konstantin~Yu. Bliokh, Yury~P. Bliokh, Sergey Savel'ev, and Franco Nori.
\newblock Semiclassical dynamics of electron wave packet states with phase
  vortices.
\newblock {\em Phys. Rev. Lett.}, 99:190404, 2007.

\bibitem{Uchida_Nature_464_737_2010}
Masaya Uchida and Akira Tonomura.
\newblock Generation of electron beams carrying orbital angular momentum.
\newblock {\em Nature}, 464(7289):737--739, 2010.

\bibitem{Verbeeck_Nature_467_301_2010}
J.~Verbeeck, H.~Tian, and P.~Schattschneider.
\newblock Production and application of electron vortex beams.
\newblock {\em Nature}, 467(7313):301--304, 2010.

\bibitem{McMorran_Science_331_192_2011}
Benjamin~J. McMorran, Amit Agrawal, Ian~M. Anderson, Andrew~A. Herzing,
  Henri~J. Lezec, Jabez~J. McClelland, and John Unguris.
\newblock Electron vortex beams with high quanta of orbital angular momentum.
\newblock {\em Science}, 331(6014):192--195, 2011.

\bibitem{Karimi:PRL:108:044801:2012}
Ebrahim Karimi, Lorenzo Marrucci, Vincenzo Grillo, and Enrico Santamato.
\newblock Spin-to-orbital angular momentum conversion and spin-polarization
  filtering in electron beams.
\newblock {\em Phys. Rev. Lett.}, 108:044801, Jan 2012.

\bibitem{Schattschneider:PRL:109:084801:2012}
P.~Schattschneider, M.~St\"oger-Pollach, and J.~Verbeeck.
\newblock Novel vortex generator and mode converter for electron beams.
\newblock {\em Phys. Rev. Lett.}, 109:084801, Aug 2012.

\bibitem{Clark:PRL_111_064801:2013}
L.~Clark, A.~B\'ech\'e, G.~Guzzinati, A.~Lubk, M.~Mazilu, R.~Van~Boxem, and
  J.~Verbeeck.
\newblock Exploiting lens aberrations to create electron-vortex beams.
\newblock {\em Phys. Rev. Lett.}, 111:064801, Aug 2013.

\bibitem{Verbeeck:APL:99:203109:2011}
J.~Verbeeck, P.~Schattschneider, S.~Lazar, M.~St\"oger-Pollach, S.~L\"offler,
  A.~Steiger-Thirsfeld, and G.~Van~Tendeloo.
\newblock Atomic scale electron vortices for nanoresearch.
\newblock {\em Appl. Phys. Lett.}, 99(20):203109, 2011.

\bibitem{Bliokh:PRL_107_174802_2011}
Konstantin~Y. Bliokh, Mark~R. Dennis, and Franco Nori.
\newblock Relativistic electron vortex beams: Angular momentum and spin-orbit
  interaction.
\newblock {\em Phys. Rev. Lett.}, 107:174802, Oct 2011.

\bibitem{Lloyd_PRL_108_074802_2012}
Sophia Lloyd, Mohamed Babiker, and Jun Yuan.
\newblock Quantized orbital angular momentum transfer and magnetic dichroism in
  the interaction of electron vortices with matter.
\newblock {\em Phys. Rev. Lett.}, 108:074802, 2012.

\bibitem{Yuan:PRA:88:031801:2013}
J.~Yuan, S.~M. Lloyd, and M.~Babiker.
\newblock Chiral-specific electron-vortex-beam spectroscopy.
\newblock {\em Phys. Rev. A}, 88:031801, Sep 2013.

\bibitem{Babiker_PhysRevLett_89_143601_2002}
M.~Babiker, C.~R. Bennett, D.~L. Andrews, and L.~C. D\'avila~Romero.
\newblock Orbital angular momentum exchange in the interaction of twisted light
  with molecules.
\newblock {\em Phys. Rev. Lett.}, 89:143601, 2002.

\bibitem{Araoka_PhysRevA_71_055401_2005}
F.~Araoka, T.~Verbiest, K.~Clays, and A.~Persoons.
\newblock Interactions of twisted light with chiral molecules: An experimental
  investigation.
\newblock {\em Phys. Rev. A}, 71:055401, 2005.

\bibitem{Loeffler_PhysRevA_83_065801_2011}
W.~L\"offler, D.~J. Broer, and J.~P. Woerdman.
\newblock Circular dichroism of cholesteric polymers and the orbital angular
  momentum of light.
\newblock {\em Phys. Rev. A}, 83:065801, 2011.

\bibitem{Boyd:NonlinearOptics:Book:2008}
R.~W. Boyd.
\newblock {\em Nonlinear Optics}.
\newblock Elsevier, 2008.

\bibitem{Pan:PhysRevB:39:1229:1989}
Ru-Pin Pan, H.~D. Wei, and Y.~R. Shen.
\newblock Optical second-harmonic generation from magnetized surfaces.
\newblock {\em Phys. Rev. B}, 39:1229--1234, 1989.

\bibitem{Pisarev:JPhysCondMat:5:8621:1993}
R.~V. Pisarev, B.~B. Krichevtsov, V.~N. Gridnev, V.~P. Klin, D.~Fr\"ohlich, and
  Ch. Pahlke-Lerch.
\newblock Optical second-harmonic generation in magnetic garnet thin films.
\newblock {\em J Phys. Cond Mat.}, 5:8621--8628, 1993.

\bibitem{Pavlov:PhysRevLett:78:2004:1997}
V.~V. Pavlov, R.~V. Pisarev, A.~Kirilyuk, and Th. Rasing.
\newblock Observation of a transversal nonlinear magneto-optical effect in thin
  magnetic garnet films.
\newblock {\em Phys. Rev. Lett.}, 78:2004--2007, 1997.

\bibitem{Kirilyuk:PhysRevB:61:R3796:2000}
A.~Kirilyuk, V.~V. Pavlov, R.~V. Pisarev, and Th. Rasing.
\newblock Asymmetry of second harmonic generation in magnetic thin films under
  circular optical excitation.
\newblock {\em Phys. Rev. B}, 61:R3796--R3799, 2000.

\bibitem{Gridnev:PhysRevB:63:184407:2001}
V.~N. Gridnev, V.~V. Pavlov, R.~V. Pisarev, A.~Kirilyuk, and Th. Rasing.
\newblock Second harmonic generation in anisotropic magnetic films.
\newblock {\em Phys. Rev. B}, 63:184407, 2001.

\bibitem{Hansteen:PhysRevB70:094408:2004}
Fredrik Hansteen, Ola Hunderi, Tom~Henning Johansen, Andrei Kirilyuk, and Theo
  Rasing.
\newblock Selective surface/interface characterization of thin garnet films by
  magnetization-induced second-harmonic generation.
\newblock {\em Phys. Rev. B}, 70:094408, 2004.

\bibitem{Koeksal:PhysRevA:86:063812:2012}
K.~K\"oksal and J.~Berakdar.
\newblock Charge-current generation in atomic systems induced by optical
  vortices.
\newblock {\em Phys. Rev. A}, 86:063812, 2012.

\bibitem{Quinteiro_OptExpr_19_26733_2011}
G.~F. Quinteiro, P.~I. Tamborenea, and J.~Berakdar.
\newblock Orbital and spin dynamics of intraband electrons in quantum rings
  driven by twisted light.
\newblock {\em Opt. Express}, 19(27):26733--26741, 2011.

\bibitem{Hansteen:PhysRevLett:95:047402:2005}
Fredrik Hansteen, Alexey Kimel, Andrei Kirilyuk, and Theo Rasing.
\newblock Femtosecond photomagnetic switching of spins in ferrimagnetic garnet
  films.
\newblock {\em Phys. Rev. Lett.}, 95:047402, 2005.

\bibitem{Bass:PhysRevLett:9:446:1962}
M.~Bass, P.~A. Franken, J.~F. Ward, and G.~Weinreich.
\newblock Optical rectification.
\newblock {\em Phys. Rev. Lett.}, 9:446--448, 1962.

\bibitem{Landau:ZdS:8:p153:1935}
L.~Landau and E.~Lifshitz.
\newblock On the theory of the dispersion of magnetic permeability in
  ferromagnetic bodies.
\newblock {\em Zeitschr. d. Sowj.}, 8:153, 1935.

\bibitem{Gilbert:ITOM:40:p3443:2004}
T.~L. Gilbert.
\newblock A phenomenological theory of damping in ferromagnetic materials.
\newblock {\em IEEE Trans. Magn.}, 40:3443, 2004.

\bibitem{Thakur_OptExpr_18_27691_2010}
Anita Thakur and Jamal Berakdar.
\newblock Self-focusing and defocusing of twisted light in non-linear media.
\newblock {\em Opt. Express}, 18(26):27691--27696, 2010.

\bibitem{Sukhov_PRL_102_057204_2009}
A.~Sukhov and J.~Berakdar.
\newblock Local control of ultrafast dynamics of magnetic nanoparticles.
\newblock {\em Phys. Rev. Lett.}, 102:057204, 2009.

\bibitem{Sukhov_PRB_79_134433_2009}
A.~Sukhov and J.~Berakdar.
\newblock Steering magnetization dynamics of nanoparticles with ultrashort
  pulses.
\newblock {\em Phys. Rev. B}, 79:134433, 2009.

\bibitem{Sukhov_AppPhysA_98_837_2010}
A.~Sukhov and J.~Berakdar.
\newblock Influence of field orientation on the magnetization dynamics
  of nanoparticles.
\newblock {\em Appl. Phys. A}, 98(4):837--842, 2010.

\end{thebibliography}
\bibliographystyle{unsrt}

\end{document}